# Evolution of the protolunar disk: dynamics, cooling timescale and implantation of volatiles onto the Earth

Sébastien CHARNOZ[1,2,3]

Chloé MICHAUT[1]

(1) Institut de Physique du Globe/ Université Paris Diderot, Paris, France

(2) Laboratoire AIM, Université Paris Diderot / CEA /CNRS, Gif-sur-Yvette, France

(3) Corresponding author: charnoz@cea.fr

# Abstract:

It is thought that the Moon accreted from the protolunar disk that was assembled after the last giant impact on Earth. Due to its high temperature, the protolunar disk may act as a thermochemical reactor in which the material is processed before being incorporated into the Moon. Outstanding issues like devolatilisation and istotopic evolution are tied to the disk evolution, however its lifetime, dynamics and thermodynamics are unknown. Here, we numerically explore the long term viscous evolution of the protolunar disk using a one dimensional model where the different phases (vapor and condensed) are vertically stratified. Viscous heating, radiative cooling, phase transitions and gravitational instability are accounted for whereas Moon's accretion is not considered for the moment. The viscosity of the gas, liquid and solid phases dictates the disk evolution. We find that (1) the vapor condenses into liquid in ~10 years, (2) a large fraction of the disk mass flows inward forming a hot and compact liquid disk between 1 and 1.7 Earth's radii, a region where the liquid is gravitationally stable and can accumulate, (3) the disk finally solidifies in $10^3$ to $10^5$ years. Viscous heating is never balanced by radiative cooling. If the vapor phase is abnormally viscous, due to magneto-rotational instability for instance, most of the disk volatile components are transported to Earth leaving a disk enriched in refractory elements. This opens a way to form a volatile-depleted Moon and would suggest that the missing Moon's volatiles are buried today into the Earth. The disk cooling timescale may be long enough to allow for planet/disk isotopic equilibration. However large uncertainties on the disk physics remain because of the complexity of its multi-phased structure.

# 1. Introduction

The Earth's Moon is believed to have formed in the aftermath of the last giant impact on the proto-earth. Whereas numerous works treating the giant-impact have been published, the subsequent evolution of the protolunar disk is only poorly known. The main challenge is to couple in a single framework both the dynamical and thermodynamical evolution of the disk material. Kokubo & Ida (2000) have simulated the re-accretion of a protolunar disk into a proto-moon from solid particles and neglected thermodynamics. The system was found to collapse into a disk and form a satellite at the Roche Limit in ~100 orbits. Once the particulate disk is completely cold, if it contains about 1% of the mass of the Earth, gravitational instabilities lead to the formation of a single moon only (Kokubo and Ida 2000; Crida & Charnoz, 2012). Machida & Abe (2004) have studied the evolution of a two-phase disk but neglected cooling and vapor condensation as well as time evolution. Salmon and Canup (2012) have studied the accretion of the Moon from a Roche interior fluid disk assuming that the disk acts as a reservoir with a radially constant surface density; its outflow is prescribed assuming that viscous heating perfectly balances radiative cooling as suggested in Thompson and Stevenson (1988). The vertical structure of a vapor/liquid magma disk at thermodynamical equilibrium is described by Ward (2012) and its evolution investigated analytically in Ward (2014a) assuming several simplifying assumptions such as a steady-state radial flux. Hence, in all previously mentioned work, the coupling between the disk dynamics and thermal evolution (radiative cooling, viscous heating and phase transitions) is not treated in a time-evolving disk. Impact simulations show that the disk is initially very hot (> 3000 K) and, in order to become gravitationally unstable and assemble into a moon, it must cool down first. During this cooling phase, a major restructuration may be expected.

The cooling timescale of the protolunar disk is actively debated as it determines the period during which the lunar material can be processed chemically as well as isotopically, before being incorporated into the Moon (Pahlevan and Stevenson 2007). A simple computation of the cooling timescale (internal energy divided by black body emission power) that ignores any dynamical evolution gives a timescale of a few 10 to a few 100 years. However, as noted by Thompson and Stevenson, (1988) and Ward (2011, 2012), if the disk is gravitationally unstable it should heat up rapidly due to a large effective viscosity induced by gravitational instabilities and this heating should increase the cooling timescale. It is even possible that viscous heat production may, at some point, perfectly balance the energy loss due to black-body emission, lengthening by orders of magnitude the cooling timescale of the disk (Thompson and Stevenson, 1988, Ward 2012). However Ward (2014a) seems to reach an opposite conclusion based on steady state models of disks. So, depending on the study, the cooling timescale may extend from ~10 years to several thousands of years.

These uncertainties have important consequences on the subsequent composition of the lunar material. Many measurements suggest a significant processing of the disk material before being incorporated into the Moon. On a three-isotope plot ($\delta^{17}O$ versus $\delta^{18}O$), different samples from different planetary bodies are all aligned on a unique fractionation line, characteristic of the body, with a slope close to 0.5. Thus, if the Moon is mainly made of impactor's material, as it is the case in the so-called canonical impact (Canup 2004), it may leave its imprint on the protolunar disk and the Moon (see e.g. Canup 2004, Pahlevan & Stevenson 2007). However, the oxygen fractionation line of the Moon is either indistinguishable from the terrestrial line (Wiechert et al, 2001) or, at most, differs by a very small fraction (Herwartz et al., 2014). In an attempt to reconcile the canonical impact model with the identical oxygen isotopic composition of the Earth and the Moon, Pahlevan and Stevenson (2007) suggested that turbulence in the disk may allow isotopic equilibration of the disk with the proto-Earth and erase any isotopic difference. The Moon is also substantially depleted in volatile elements compared to the Earth's mantle, by a factor of 10 for moderately volatile elements like potassium and almost by a factor of 100 for highly volatile elements like Zinc (Wieczoreck & Taylor 2014). In addition Zinc isotopes are strongly enriched in their most heavy species (Paniello et al., 2012) arguing for an efficient volatile removing process. An incomplete condensation of the Moon was suggested by Stewart et al (2013) but hydrodynamical escape of volatile elements may not be efficient enough to explain Moon's depletion because of the high concentration of oxygen atoms in the disk and possibly its short lifetime (Nakajima & Stevenson 2014b). The processes by which the lunar material has become depleted in volatile elements are thus still questioned.

In order to make progress in our understanding of the disk, the present work aims to build a numerical model of a time-evolving protolunar disk.

In their pioneering work, Thompson and Stevenson (1988) described the complex physical processes at play in the protolunar disk making it particularly exotic compared to many other astrophysical disks. The disk is likely to be constituted by a two-phase media, with a liquid and a vapor phase. The disk cools from its upper atmosphere and generates heat through viscous dissipation in the midplane. The vapor phase condenses in liquid droplets that sediment in the midplane. Whereas Thompson and Stevenson (1988) assumed that the disk is made of an intimate mixture of gas and droplets, more recent works suggest that the disk is indeed stratified with a condensed layer in the midplane (liquid or solid) topped by a vapor atmosphere. Machida & Abe (2004) have shown that the droplets sediments in less than $10^{-2}$ years. Recently, Nakajima and Stevenson (2014a) proposed scaling relations for estimating the disk vapor fraction in good agreement with SPH simulations, assuming such a vertically stratified disk at hydrostatic equilibrium.

Unfortunately there is no numerical code today that can grasp all the physical ingredients of the protolunar disk as described in Thompson & Stevenson (1988) and compute the disk evolution over a cooling timescale (> 100 years).

Here we propose a model for a two-phase, vertically stratified protolunar disk. Our central idea is to use technics developed for the study of protoplanetary disks and adapt them to the case of the protolunar disk to account in particular for the presence of two phases, which imposes major changes compared to all published protoplanetary disk models. We numerically track the evolution of the protolunar disk, just after the giant impact, and over hundreds of thousands of years with a simple, but non-trivial, one dimensional and two-phase model. On the contrary to Machida and Abe (2004), we allow for mass exchanges between the vapor and liquid phases (due to cooling and condensation), consider a time evolving temperature and account for radiative cooling and viscous heating. We emphasize the role of viscosity as the main driver of the disk evolution over long timescales (i.e. timescales >> orbital period which is ~10 hours on average), and of phase transition effects. To make the computation tractable, we assume that the disk is always in hydrostatic equilibrium and locally vertically isothermal. The aim of the present study is to focus on the disk evolution. The growing of a proto-moon and its back-reaction onto the disk are not considered. Taking this process into account may imply significant modifications of our (already complex) code. In order to precisely understand the effects of each process on the disk evolution, we choose a step-by-step approach. We leave the effect of a proto-moon for a future work (whereas a simplified case is presented in section 3.3.3 and in section 3 of the supplementary online material).

The paper is organized as follows: in section 2 the physics and the algorithm that couple the dynamical evolution, thermodynamical equilibrium, heating and cooling are described. A special care is given to the computation of viscosities that are key ingredients. In section 3 we present the evolution of a disk, first neglecting the disk's dynamical evolution (in order to illustrate the basic effects of the cooling and heating processes involved) and then considering the disk full thermodynamical and dynamical evolution. Cases with fully separated phases and with some degrees of mixing are compared. In section 4, the evolution of the disk content in volatile and refractory species is presented and we show that it is possible to devolatilize the disk rapidly if the vapor layer is abnormally viscous. In section 5 the results are commented in the context of different Moon formation scenarios. In section 6 we summarize our results and conclude on their possible implications regarding Moon's formation.

## 2. The model

### 2.1 Overall structure of the model

We solve for the one dimensional time evolution of the condensed phase (liquid or solid) and the vapor phase, in terms of surface density and temperature as a function of distance to the proto-Earth. This approach is inspired from one-dimensional codes developed for the study of protoplanetary disks or rings (see e.g. Hueso and Guillot 2008, Charnoz et al., 2011, Yang and Ciesla 2012, Baillie and Charnoz, 2014), the main difference being the taking into account of two phases (Figure 1). For simplicity, the two layers of the disk are assumed to be in local hydrostatic equilibrium and the disk is assumed to be vertically isothermal. In fact, the vapor layer is probably everywhere in equilibrium with the condensed phase and should lie on the Clapeyron curve (Ward 2012). But the structure and energy budget of an isothermal vapor layer is very close to the case of a Clapeyron atmosphere (see Annexe A of the present paper, Genda & Abe 2003; Ward 2012). Another major simplification is that the disk is assumed to be at centrifugal equilibrium at all times, so that the sum of the pressure gradient (divided by density) and the Earth's gravity always perfectly balances the centrifugal force. In other terms, we do not consider the propagation of pressure waves. These waves travel rapidly accross the disk, imposing too small time-steps and preventing the investigation of the long-term evolution of the system. Thus, effects occurring on timescales smaller than the sound propagation timescale (that is about 2 to 5 hours to travel on the disk radius) are ignored.

A time-step is divided in three phases:

1) **Dynamical evolution**: the surface densities of the vapor and condensed layers evolve according to the continuity and conservation of angular momentum equation (including the viscous term). Mass and Internal energy are transported according to the local velocity field.

2) **Radiative cooling and viscous heating:** the two layers exchange and loose internal energy through their own black-body emission, and produce internal energy through viscous dissipation. This step modifies the internal energy (and, thus, the temperature) of each phase.

3) **Heat exchange and mass exchange due to phase transition:** The two layers exchange mass and energy as they return to thermal equilibrium so that they always end with the same temperature.

We now describe in details each of these steps.

### 2.2 Dynamical evolution

Let $\sigma$ and $U$ be the surface density and radial velocity of a given phase (condensed or vapor). Subscripts *l* and *v* respectively designate the condensed phase (liquid or solid) and the vapor phase. The evolution of the surface density obeys the mass conservation equation:

$$\frac{d\sigma_l}{dt} + \frac{1}{r}\frac{d}{dr}(\sigma_l U_l r) = 0$$
$$\frac{d\sigma_v}{dt} + \frac{1}{r}\frac{d}{dr}(\sigma_v U_v r) = 0$$
**Equation 1**

Where the radial velocities of liquid or solid $U_l$ and vapor $U_v$ are given by the conservation of angular momentum taking into account the viscous term (see e.g. Takeuchi et al., 1996) :

$$U_l = \frac{\frac{d}{dr}\left[r^3 \nu_l \sigma_l \frac{d\omega_l}{dr}\right]}{\frac{dr^2\omega_l}{dr}}$$

$$U_v = \frac{\frac{d}{dr}\left[r^3 \nu_v \sigma_v \frac{d\omega_v}{dr}\right]}{\frac{dr^2\omega_v}{dr}}$$
Equation 2

where ω is the angular velocity. We assume that $\omega_l = \omega_k$, where $\omega_k$ is the keplerian orbital frequency (Maschida & Abe, 2004) owing to the large density of liquids (so that the pressure support is weak). For the vapor, the orbital frequency is subkeplerian because of pressure support. The difference between the orbital frequency of the vapor ($\omega_v$) and the keplerian frequency is comparable to the thermal kinetic energy of the vapor divided by the kinetic energy of the orbital velocity: in other words, $\omega_v \sim \omega_k(1-(C_{sv}/r\omega_k)^2)$ with $C_{sv}$ standing for the local sound velocity in the vapor ($C_{sv} = \left(\frac{\gamma RT}{\mu}\right)^{1/2}$, with µ=30 g/mol standing for the average molar mass of vaporized rock (Ward, 2011), γ=5/3 the adiabatic exponent, R the ideal gas constant). The difference between $\omega_v$ and $\omega_k$ is only a few percents (Machida & Abe 2004).

**2.3 radiative and viscous heat production evolution**

Let *e* denotes the internal energy per surface area. The internal energy transport equation is for the liquid and vapor:

$$\frac{d\,e_l}{dt} + \frac{1}{r}\frac{d}{dr}(e_l U_l r) = \dot{e}_{lV} + \dot{e}_{lR}$$

$$\frac{d\,e_v}{dt} + \frac{1}{r}\frac{d}{dr}(e_v U_v r) = \dot{e}_{vV} + \dot{e}_{vR}$$
Equation 3

Where the compressibility term (Pdv) is not considered as we assume that adiabatic compression is negligible compared to radiative cooling and viscous heating, as in the case of protoplanetary disks.

The terms $\dot{e}_{lV}$, $\dot{e}_{lR}$, $\dot{e}_{vV}$, $\dot{e}_{vR}$ stand for energy production and loss in the liquid and vapor by viscous dissipation (subscript V) and radiative processes (subscript R). The viscous dissipation terms are as follows (Pringle, 1981):

$$\dot{e}_{lV} = +\frac{1}{2}\sigma_l \nu_l r^2 \left(\frac{d\omega_l}{dr}\right)^2$$

$$\dot{e}_{vV} = +\frac{1}{2}\sigma_v \nu_v r^2 \left(\frac{d\omega_v}{dr}\right)^2$$
Equation 4

meaning that the liquid layer and the vapor layer produce energy through viscous dissipation. The radiative cooling terms are given by:

$$\dot{e}_{lR} = -2\sigma_S T_l^{\,4} + 2\sigma_S T_v^{\,4}$$

$$\dot{e}_{vR} = -2\sigma_S T_v^{\,4} + 2\sigma_S T_l^{\,4} - 2\sigma_S T_{irr}^{\,4}$$
Equation 5

Where $T_v$ and $T_l$ are the temperatures of the vapor and condensed phase (liquid or solid) respectively. Equation 5 means that the liquid disk loses energy from its surface due to black body emission, but also that it absorbs the vapor disk black-body emission. The gas layers have two interfaces with the liquid disk, where radiative energy is exchanged with the liquid disk at the current local temperature, and two free interfaces where energy is ultimately radiated away to space with an effective temperature $T_{irr}$ that corresponds to the condensation temperature of silicates (~ 2000 K). The above equations are somewhat similar to those of Bell & Lin (1994) though they do not consider irradiation from the central body (which could be a future improvement).

### 2.4 Mass and energy exchange due to phase transition

After the dynamical and energy steps, the temperatures of the liquid and vapor layers are different. In the last step, phase transition is computed so that thermal equilibrium is restored and the liquid and vapor disks locally return to the same temperature by material exchange. For a given total surface density (liquid plus vapor) and a given temperature, only a specific quantity of mass can be stored in the vapor layer in order to exert enough pressure at its base to embed a liquid midplane layer. So the mass stored in the vapor and condensed phases must be computed first.

#### 2.4.1 Hydrostatic equilibrium of the vapor and liquid layers

Let $T_e$ be the temperature of the vapor and liquid layers at thermodynamical equilibrium. At the interface between the two layers, the vapor is in equilibrium with the liquid (on the Clapeyron curve) so that there is a simple relation between pressure ($P_i$) and temperature at the liquid/vapor interface:

$$P_i(T_e) = P_0 e^{-T_0/T_e} \qquad \textbf{Equation 6}$$

Where $P_0$ and $T_0$ depend on material properties. As in Ward (2012) we adopt the following values for silicate vapor/liquid equilibrium : $P_0= 3\times10^{13}$ Pa, $T_0=60000$K. Hydrostatic and isothermal equilibrium of the vapor layer implies that the vertical distribution of density is (Takeuchi & Lin 2002):

$$\rho_g(z) = \rho_{g,0} e^{-\frac{z^2}{2H^2}} \qquad \textbf{Equation 7}$$

With H standing for the local pressure scale-height $H = \frac{C_{sv}^2}{\omega_k}$ (Takeuchi & Lin 2002). Assuming that the base of the vapor layer is in equilibrium at temperature $T_e$ with the underlying liquid layer, we combine Equation 6 and Equation 7 so that $\rho_g(z=0) C_{sv}^2(T_e) = P_i(T_e)$. Then, if we assume that the thickness of the condensed layer is negligible compared to that of the vapor layer, the surface density $\sigma_v$ is the integral of the density from -∞ to +∞ so that $\sigma_v(r)=(2\pi)^{1/2} H \rho_g(z=0)$ (Takeuchi & Lin 2002) and we get the surface density of the vapor layer at temperature $T_e$. Noting $\sigma_{ve}(T_e)$ the surface density of vapor at thermodynamical equilibrium:

$$\sigma_{ve}(T_e) = \frac{(2\pi)^{1/2} P_i(T_e) C_{sv}(T_e)}{\omega_k} \qquad \textbf{Equation 8}$$

The mass fraction of vapor, f(T), at temperature $T_e$ is $f(T_e)=\sigma_{ve}(T_e)/\sigma$ where $\sigma=\sigma_v+\sigma_l$ is the total local surface density. So, at temperature $T_e$, only the amount of vapor given by Equation 8 will be stored in the vapor layer. The remaining material makes the condensed midplane layer. The local thermodynamical equilibrium (see the following section) determines the temperature $T_e$ (see the following section)

#### 2.4.2 Solving for the local thermodynamical equilibrium

After the dynamical evolution step and the radiative cooling/viscous heating step, the vapor and condensed layers have densities $\sigma_l$ and $\sigma_v$, temperatures $T_l$ and $T_v$ and internal energies (per unit surface) $e_v=C_{vv} \sigma_v T_v$ and $e_l=C_{vl} T_l \sigma_l$, where $C_v$ is the heat capacity at constant volume. For the vapor, the heat capacity is given by $C_{vv} = \frac{R}{\mu(\gamma-1)}$, and, for the condensed phase, we use $C_{vl}=416\times10^3$ J/kg as a fiducial value (Ward, 2012). We compute the surface densities of the vapor and liquid layers at equilibrium, $\sigma_{le}$ and $\sigma_{ve}$, and the equilibrium temperature $T_e$ using an iterative procedure such that the energy given by the vapor layer is equal to the energy received by the liquid layer. For a vapor layer the total energy is the sum of its enthalpy ( $(C_{vv}+R/\mu)\,\sigma_v\,T_v$ ) and its potential energy above the midplane (1/2 $\sigma_v\,C_{sv}^2$) . The latent heat given to the liquid layer by the liquefaction of a surface mass density ($\sigma_{le}$-$\sigma_l$) is $l(\sigma_{le}$-$\sigma_l)$ where $l$ is the latent heat of liquefaction. When the two layers come back to thermodynamical equilibrium at temperature $T_e$, the vapor layers goes from temperature $T_v$ to temperature $T_e$ while the liquid layer goes from $T_l$ to $T_e$. During this transformation, the energy received by the liquid layer is $\sigma_l C_{vl}(T_e - T_v)$ so that the energy balance reads: (energy given by the vapor = energy received by the liquid):

$$\sigma_v \left(C_{vv} + \frac{R}{\mu}\right)(T_v - T_e) + \frac{1}{2}\sigma_v\bigl(C_{sv}^2(T_v) - C_{sv}^2(T_e)\bigr) + l(\sigma_{le} - \sigma_l) = \sigma_l C_{vl}(T_e - T_v)$$

**Equation 9**

And to ensure mass conservation:

$$\sigma = \sigma_{ve}(T) + \sigma_{le}(T) \qquad \text{Equation 10}$$

To find the final state at equilibrium, i.e. $\sigma_{ve}$, $\sigma_{le}$, and $T_e$, Equations 8, 9 10 must be solved simultaneously for $T_e$. A Newton-Raphson method is used to solve this system at each time step and at every location. At the end of the time step, $\sigma_{ve}$, $\sigma_{le}$ are the new values for the local surface densities of vapor and liquid and the new values of internal energies are $e_v=C_{vv}\,\sigma_{ve}\,T_e$ and $e_l=C_{vl}\,T_e\,\sigma_{le}$ for the vapor and condensed layers respectively.

### 2.5 Viscosity

The local viscosity is a critical parameter as it induces both transport (section 2.2) and viscous heating (section 2.3). Unfortunately it is poorly known. In addition, because of the possible onset of Kelvin-Helmotz (KH) instability (Thompson & Stevenson 1988, Sekiya 1998, Machida & Abe 2004), there is a possibility for the condensed phase to be mixed with the vapor phase to some extent. Beside the molecular viscosity, anomalous sources of viscosity may be important such as in the case of many astrophysical contexts. In the context of the protolunar disk two possible sources of anomalous viscosity are obvious: gravitational instability, which is relevant for the condensed phase either liquid or solid, and turbulence driven by a magnetic field, relevant for the ionized gas phase (although turbulence is not systematically a source of anomalous viscosity, like the thermal turbulence, see Pahlevan & Stevenson 2007 for example). We detail below how the viscosity is computed for the different phases.

### 2.5 Viscosity of the condensed phase

We assume that the solidification temperature of silicates is independent of pressure and is $T_s$=1300 K. When the disk material is solid, it is expected to be mostly granular.

### 2.5.1 Viscosity of the solid condensed phase

We take advantages of the numerous studies published on the viscosity of Saturn's rings that provides calibrated and numerically tested expressions of the effective viscosity of a granular media. We assume that the "grains" have an average size of $r_g$=1m. This value is arbitrarily chosen as we have no clue about the average size of condensates in the protolunar disk. Their density is $\rho_g$= 3000kg/m$^3$ and their mass is $M_g$=4/3 $\pi\,\rho_g\,r_g^3$. The viscosity can be decomposed into three contributions: viscosity arising from physical collisions ($\nu_{coll}$), from the non-zero size of particles ($\nu_{trans}$) and gravitational interactions ($\nu_{grav}$). Following Daisaka et al. (2001) we define the Toomre parameter for a disk made of solid particles as follows:

$$Q_s(r) = \frac{C_d \omega_k}{3.36\, G \sigma_l(r)} \qquad \text{Equation 11}$$

The effective viscosity $\nu$ is computed following Daisaka et al. (2001)'s prescription:

$$\nu = \nu_{trans} + \nu_{grav} + \nu_{coll} \qquad \text{Equation 12}$$

$$\nu_{trans} = \begin{cases} \frac{C_g^2}{2\omega_k}\left(\frac{0.46\tau}{1+\tau^2}\right) & \text{if } Q_s > 2 \\ \frac{1}{2} 26\, r_h^{*\,5} \frac{G^2 \sigma_l(r)^2}{\omega_k^3} & \text{if } Q_s < 2 \end{cases} \qquad \text{Equation 13}$$

$$\nu_{coll} = r_g^2 \omega_k \tau \qquad \text{Equation 14}$$

$$\nu_{grav} = \begin{cases} 0 & \text{if } Q_s > 2 \\ \nu_{trans} & \text{if } Q_s < 2 \end{cases} \qquad \text{Equation 15}$$

where $r_h^* = r_h/2r_g$ with $r_h = r(2M_g/3M_e)^{1/3}$ and $\tau$ is the geometric optical depth (it is the total surface of particles divided by the system surface, and is about the number of collisions per particle per orbit) $\tau = 3\sigma_l/(4 r_g \rho_g)$. The value $r_h^*$ is about the ratio of the mutual Hill sphere of two grains to their diameter. Whereas theory predicts the gravitational instability to develop for $Q_s \leq 1$, numerous numerical simulations find that the instability develops as soon as $Q_s \leq 2$ and the effective viscosity strongly increases, enhancing radial transport of material and heat production. The velocity dispersion of grains $C_d$ is computed as follows (Daisaka et al., 2001): when $r_h^* < 0.5$, the Hill sphere of the grains is smaller than their physical diameter so that encounters are dominated by rebound at the particles surface and hence $C_d = 2\omega_k r_g$. Conversely for $r_h^* \geq 0.5$ encounters among grains are mostly gravitational (as the grains are physically smaller than their mutual Hill Sphere) and thus $C_d$ is comparable to the grains escape velocity $C_d = (G M_g/r_g)^{1/2}$ (Salo 1995, Daisaka and Ida 1999, Ohtsuki 1999, Salmon et al., 2010). In most cases encountered in the present paper, the viscosity of the solid phase is dominated by the self-gravity contribution (Equation 15), so that the effective viscosity of the solid phase is high in the largest portion of the disk. It may get low close to the planet where the disk shear is strong and where the particle incompressibility may act against the gravitational instability (Takeda and Ida 2001). Close to the planet the collisional viscosity dominates (due to collisions between particles) whereas gravitational viscosity dominates far from the planet.

### 2.5.2 Viscosity in the liquid condensed phase

When the condensed phase is liquid ($T > T_s$), the situation is much less known. We adopt below the approach of Machida & Abe (2004). Theoretical studies in astrophysical incompressible fluids (Sekiya, 1983, 1998) find a unique criterion for the onset of the gravitational instability in a keplerian disk: the disk gets gravitationally unstable when the distance to the planet r becomes larger than $R_L$ with

$$R_L = \left(\frac{M_e}{0.525 \pi \rho_l}\right)^{1/3} \qquad \text{Equation 16}$$

where $\rho_l$ stands for the material density of the liquid (we use $\rho_l = 3000$ kg/m$^3$). Using these values $R_L$ = 1.7 $R_\oplus$. Thus, all fluid material is thus gravitationally unstable for $r > 1.7\, R_\oplus$ (with $R_\oplus$ standing for the Earth's radius). So, when the condensed phase is in liquid form (T>$T_s$), the following prescription for the viscosity is applied

$$\begin{cases} \nu = \dfrac{\pi^2 G^2 \sigma_l^2}{\omega_k^3} & for\ r > R_L \\ \nu = 0.001\ m^2 s^{-1} & for\ r \leq R_L \end{cases} \qquad \textbf{Equation 17}$$

For r>$R_L$ the disk is gravitationally unstable and an anomalous viscosity is considered, valid for self-gravitating disk, while interior to $R_L$, the liquid is stable and its viscosity reduces to its molecular viscosity, chosen as 0.001 $m^2s^{-1}$ as is typical of magma at high temperature on Earth. Although the molecular viscosity of magmas varies over several orders of magnitude depending on their chemical composition and the presence of dissolved water, in our case, this has little influence on the disk evolution as these viscosities are very low compared to the case when the disk is self-gravitating.

Note however that the effective viscosity of the liquid disk, due to its incompressible nature, is quite uncertain. Stewart (2000) pointed out that surface tension of the sort considered by Thompson and Stevenson (1988) can also stabilize magma droplets against tidal shear for droplets up to a good fraction of a meter in radius. This suggests that a behavior similar to a particle disk may be possible. We do not consider this possibility here and follow the same approach as Sekiya (1983) and Machida and Abe (2003) but discuss this possibility in the last section. Note however that a disk that is gravitationally unstable far from the planet and stable close to it is in agreement with findings reported in several previous studies (Ward and Cameron 1978, Daisaka and Ida 1999, Takeda and Ida 2001). For example, Ward and Cameron (1978) predicted that gravitational instability does not occur when r<0.5xRoche Limit ~ 1.8 Earth radii~$R_L$. Daisaka and Ida (1999) find a similar result considering the shrinkage of the Hill sphere close to the planet.

**2.6 Viscosity of the vapor layer**

Owing to its high temperature, the vapor phase is never prone to gravitational instability. However, it is expected to be substantially ionized (see figure 2 of Visscher & Fegley 2013) which may favor the development of the magneto-rotational instability (Balbus and Hawley 1991), MRI hereafter, provided that there is an ambient magnetic field. MRI is known to induce a strong effective viscosity implying transport of angular momentum and heat production (see e.g. Balbus and Hawley 1991, Fromang 2013, Flock et al., 2013). Thus, in this work we consider two possibilities for the vapor layer: either (1) the vapor layer has no anomalous source of viscosity or (2) it has an anomalous source of viscosity, like MRI. In the first case we assume that the gas molecular viscosity scales with the square root of temperature (like for an ideal gas) so that: $\nu_g = 10^{-3}$ (T/300K)$^{1/2}$ $m^2s^{-1}$. In the second case we assume that the vapor layer suffers MRI and its effective viscosity may be described using a popular alpha-disk model (Shakura and Sunyaev, 1973) so that:

$$\boldsymbol{\nu_g = \alpha \dfrac{C_{sv}^2}{\omega_k}} \qquad \textbf{Equation 18}$$

In the current state of our knowledge it is not known if the protolunar disk is MR stable or not, but there is no reason, a priori, to reject this possibility: owing to its high temperature, the electron fraction is very high (~$10^{-3}$ see Visscher & Fegley 2013) and the presence of a magnetic field is possible either due to a primordial magnetic field of the proto-earth or due to the local solar magnetic field. Of course $\alpha$ is an unknown parameter but numerous simulations of perfectly magnetized protoplanetary disk show that 0.0001<$\alpha$<0.1 (see e.g. Fromang & Nelson 2006). MRI may also affect material diffusivity with gas diffusion coefficients scaling with viscosity (Fromang & Nelson 2006). So $\nu_g$ quantifies also the intensity of diffusion of chemical species. In their seminal paper, Pahlevan & Stevenson 2007 suggests $\alpha$~$10^{-4}$ (but they assume thermal turbulence, not MRI). Here, we will use this value of $\alpha$ when considering viscous vapor disks. Note however that the formation of liquid droplets in the vapor layer may de-ionize the gas, acting against MRI. So the two cases considered above may be seen as two extreme cases (no MRI and fully developed MRI).

**2.7 Concluding Remarks about disk viscosities**

Before closing this section on viscosities it is important to emphasize that the viscosities of vapor, solid and liquid are fundamentally different. These differences will play an important role in the following results. When the disk is in the form of solid particles it can be viscous everywhere even *below* $R_L$, due to the collisional viscosity that can induce a high effective viscosity, even in the absence of gravitational instability close to the planet. Conversely, gravitational instability is impossible for an incompressible fluid disk below $R_L$ according to Sekiya's result (1983) (that is still surprising … at least for the authors of the present paper) resulting in a low-viscosity liquid disk close to the planet. The Sekiya's results may be recovered assuming that an incompressible fluid behaves as if it was a gas disk with sound velocity always equal to $H\omega_k$ with $H=\sigma_l/\rho_l$ (representing the geometrical thickness of a disk of incompressible material). Indeed Sekiya's 1983 critical radius for fluids, $R_L$ is found to a constant factor of order 1, by setting the sound velocity to $\sigma_l/\rho_l \, \omega_k$ and applying the classical Toomre coefficient instability criterion for a gas. Unfortunately there is almost no study of incompressible keplerian disks apart from those of Sekiya. This interesting result deserves to be investigated with an incompressible 3D numerical simulation that is well beyond the scope of the present paper.

A last question, that deserves to be put forward, is the role of these effective viscosities with respect to material heating. Viscosities imply heat generation through shear heating. As these viscosities do not represent real viscosities but rather transport coefficients for different kinds of quantities (angular momentum), there is, *a priori*, no reason why the viscosity term that appears in the angular momentum equation (Equation 2) is exactly the same as the viscosity arising in the heat generation equation (Equation 4). In the case of Magneto Rotational Instability in a protoplanetary disk, however, the link between the effective alpha-viscosity (relevant for transport of angular momentum) and the amount of heat generated, averaged over the disk thickness, has been demonstrated to be conformed to Equation 2 (see for example the beautiful work of Flock et al., 2013). For any disk, an effective viscosity for the transport of angular momentum induces a net loss of energy. Some of this energy loss is taken on potential gravitational energy (induced by the inward flow of material in average), and the remaining energy must be taken somewhere. In all published studies of protolunar disks, it is assumed that this remaining energy is taken in the form of heat that is ultimately radiated away. But in reality this is unclear: whereas it is easy to extract heat from shock-waves induced by spiral arms in a gas disk, how can heat be extracted from an incompressible liquid or a particulate disk? One may argue that two colliding spiral arms constituted of solid particles (like in Saturn's rings) will scatter particles in all directions, thus inducing random motion among particles which corresponds precisely to a heating process. Dissipative collisions among particles will lead to fracturation, deformation, heating and ultimately, radiative cooling. For a fluid, we may imagine that two colliding spiral arms may merge to lower their surface energy, and that the remaining energy serves to heat up the fluid. In the present paper, following Machida & Abe (2004) or Ward (2012) we assume that the viscosity in the angular momentum equation is the same as the viscosity appearing in the heat-generation equation, keeping in mind that this is a matter of debate.

**2.8 Putting all pieces together**

At the beginning of a time step, the system is defined with by the surface density and internal energy per unit mass $\sigma_l$ (r) and $e_l$ (r) for the condensed phase and $\sigma_v$ (r) and $e_v(r)$ for the vapor phase. Since we assume local thermodynamical equilibrium at the beginning of a time-step we always have $T_v(r)=T_l(r)$.

First we compute the dynamical evolution of the system by

1) Computing the local viscosity (section 2.5)

2) Computing the velocity field (see Equation 2)

3) Evolving the surface densities and internal energies (Equation 1 and Equation 3)

The densities and energies are evolved using a finite volume method expressed in a conservative form to ensure perfect conservation of mass and energy. A Lax-Wendroff advection algorithm is used with a Van-Lear flux-limiter for computing the numerical flux between two neighboring cells. This ends the dynamical evolution step. At this point the system is out-of thermodynamical equilibrium. Adopting an operator splitting approach (so that each physical effect is treated one after the other), we now compute the thermodynamical evolution by:

4) Computing the amount of energy generated through viscous heating and loss through black body radiations (Equations 4 and 5) and updating the values of internal energy for the two layers.

5) Computing the thermodynamical equilibrium though mass and energy exchange between the two phases which requires to solve Equations 8,9 and 10 simultaneously and find the new values of $\sigma_l$, $\sigma_v$, $e_l$, $e_v$, $T_l$ and $T_v$ (with $T_v=T_l$ for any r)

6) Go back to 1 for a new time-step.

### 2.9 Boundary conditions: Earth's Rotation

At the outer edge, we assume that material can leave the system but cannot re-enter, so that the radial velocity at the outer-most radial bin is always assumed to be greater than, or equal to, 0 (outflow boundary condition). The boundary condition at the inner edge is supposed to describe the connection of the protolunar disk with a rotating Earth. The first 5 cells of the radial grid are assumed to represent points below the Earth's surface (located at $r=R_e$). To do so for these cells the radial velocities of both layers is 0. Since the Earth is assumed to rotate rigidly, the azimuthal velocity of cells with $r < R_\oplus$ is fixed to $\omega_e r$. The rotation period of the Earth is assumed to be 4h according to canonical models for Moon's formation (see e.g. Canup 2004, 2013).

### 2.10 What about Moon's accretion?

In the current version of the code, Moon's accretion and growth is not considered because of numerical limitations. For moon accretion to set-in, a gravitationally unstable disk is a necessary, though not a sufficient condition. Indeed the disk must also fragment to give birth to individual aggregates. If fragmentation does not occur, the gravitational instability tends to manifest itself as spiral density structures only. The criterion for fragmenting a disk has been investigated for a particulate disk or a compressible disk. For the particulate disk case the disk fragments close to the so called "Roche Limit", that is $R_R=2.456 R_e(\rho_e/\rho_g)^{1/3}$ with $\rho_e$ standing for the average Earth's density (5500 kg/m$^3$) and $\rho_g$ for the particle grain density (3000 kg/m$^3$), giving $R_R=3 R_\oplus$ for our case. However, for an incompressible liquid disk, the situation is unclear: whereas the gravitational instability operates for $r>R_L$ (=1.7 $R_\oplus$, see section 2.6.2) it is not clear if the liquid material can fragment into a single object so close to the central body (as assumed in Machida & Abe 2004), this is almost 2 times closer to the planet than for a solid particle disk. Since the three phases may fragment at different locations in the disk, the question of where the disk may fragment and give birth to a proto-Moon is complex and deserves a dedicated study. The present study is limited to the long-term evolution of the disk without considering Moon's formation. However, in order to have a first, but incomplete, insight on the effects of a growing proto-moon on the disk evolution, we present the results of one simulation with a sharp edge at the Roche Limit, that mimics roughly the first order effect of a growing proto-moon on the disk (see section 3.3.3 and supplementary online material). Note that this approach is not physically consistent since a real consideration of a growing proto-moon would imply to compute the Moon's torque onto the disk at every mean motion resonance as well as to compute the Moon's orbital evolution. This would add considerable complexification of the present model. Adopting a step-by-step approach, we focus on the disk's physics in the present paper and reserve the effects of a growing proto-moon to a future study.

## 3. Results: large scale and long term evolution of the protolunar disk

### 3.1 initial conditions

Our initial conditions are intended to represent the protolunar disk after the impact, when the disk has relaxed enough so that it is centrifugally supported. We use 100 radial cells, linearly spaced between $0.7 R_e$ and $10 R_e$. We start with the following surface density profile:

$$\sigma(r) = 4.5 \, 10^8 \left(\frac{r}{R_e}\right)^{-3} \text{ kg/m}^2 \qquad \text{Equation 19}$$

This profile is consistent with a disk resulting from a mars-sized object impacting the Earth and producing a disk with ~1.5 lunar masses ($M_m$) in total, with 0.75 $M_m$ below the classical Roche Limit (2.9 $R_e$). We assume that the disk is isentropic after the impact (Nakajima and Stevenson 2014a), resulting in a radially constant vapor mass-fraction $f=\sigma_v(r)/\sigma(r)$. Two limiting cases will be explored: f=20% and f=80%. Setting f=20% is relevant for low kinetic energy impact models where the impactor is a Mars-like object (Canup, 2004, Nakajima & Stevenson 2014a) whereas high initial vapor fraction are more relevant to scenarios with more initial kinetic energy (like in the sub-earths case, Canup 2012).

### 3.2 Thermodynamical evolution of a disk with no dynamics

In order to understand the basic physical processes occurring in the protolunar disk we present, as a pedagogical case, the time evolution of a disk with a non-viscous vapor (apart from the molecular viscosity), an initial vapor mass fraction f=20% and artificially imposing a zero radial velocity. We observe that just outside $R_L$ (1.7 Earth radii, see section 2.6.1) and up to about $4R_\oplus$ the disk is dominated by viscous power production (Figure 2 column 3). In the rest of the disk (beyond 4 $R_\oplus$) efficient cooling takes place, inducing a rapid condensation of vapor into liquid (Figure 2, column 1). A steady state is reached in about ~200 years (Figure 2, column 3, rows 2 and 3). The vapor layer has now efficiently condensed in liquid. Since liquid is more viscous that vapor (section 2.6.2), viscous heat generation increases and now perfectly balances radiative cooling below 4 $R_e$. Beyond 4 $R_\oplus$ the disk is dominated by radiation loss. However, as soon as the disk temperature drops below the solidification temperature ($T_s$=1300 K here), the viscosity suddenly jumps again because of the high viscosity of granular material (see section 2.6.1). This sudden viscosity increase produces spikes in heat production (Figure 2 column 3, rows 3 and 4). As a result, the disk stabilizes around the solidification temperature ($T_s$). At this point, the disk has reached a thermodynamical steady state beyond $R_L$ and does not evolve much. Inside $R_L$, since the liquid disk is gravitationally stable, its viscosity is low and the viscous heat generated is much lower than radiative loss. As a result the disk cools down inside $R_L$. The balance between viscous heating and radiative cooling, as observed here in the gravitationally unstable liquid disk between 1.7Re and 4Re, has been anticipated or considered in several papers (see e.g. Thompson & Stevenson, 1988; Ward 2012; Salmon & Canup, 2013). However, in the present example, the disk dynamical evolution is artificially shut off, so that the disk cannot lose potential energy through inward transport of material. As a consequence, the disk cannot completely get rid of its thermal energy. Thus its cooling timescale is infinite (when neglecting material motion). Such a balance is not realized when the fluid's dynamical evolution is considered.

**3.3 Dynamical and thermodynamical evolution of a protolunar disk with a non-viscous vapor**

In this section we present the evolution of a typical disk assuming that the vapor-phase is not abnormally viscous ($\alpha=0$), the vapor viscosity always remaining equal to the intrinsic molecular viscosity of the gas phase. Because of the very high value of the latent heat of vaporization, switching from 20 to 80 weight % (wt. % hereafter) vapor only slightly modifies the initial temperature profile, and the initial vapor fraction can almost be considered as a free initial parameter, as noted in Machida & Abe (2004).

**3.3.1 A disk with 20 wt. % vapor**

The first years of evolution are displayed in Figure 3. As the condensed phase is initially liquid, it is gravitationally unstable at all $r>R_L$ (1.7 $R_\oplus$) and thus its effective viscosity is high. So viscous heating in the liquid dominates by orders of magnitude the radiative cooling initially (Figure 3, column 3). Still the liquid is not significantly vaporized because the viscous heat production is rapidly balanced by the energy loss due to the inward flow of liquid in the region below 2.5 $R_\oplus$ (Figure 3, column 4). As a result of conservation of angular momentum the disk's outer portion flows outward. After 2 years the disk cools efficiently beyond $R_L$ and at this point, the radiative cooling overcomes viscous heating everywhere (Figure 3, row 3, column 3). After 6 years of evolution, the protolunar disk becomes mainly an accretion disk and the material flow is mostly directed inward (Figure 3, row 3, column 4). In consequence, the material accumulates below $R_L$ and the surface density strongly increases between $R_\oplus$ and $R_L$ (Figure 3, row 3, column 1). We note that most of material almost does *not* fall onto the Earth because liquid orbiting at distance below $R_L$ is gravitationally stable and, so, has a low viscosity. As a consequence the liquid material accumulates between $R_\oplus$ and 1.7 $R_\oplus$, forming a very dense hot and liquid compact disk in less than ten years, with a surface density rising up to $10^9$ kg/m$^2$. After 10 years more than 99% of the total disk mass is contained between $R_\oplus$ and $R_L$. The formation of this dense and compact inner disk is also visible in previously published studies of protolunar disk evolution, using N-body simulations with bouncing particles (but neglecting thermodynamics). For example, in Figure 4.b of Takeda and Ida (2001) shows a strong increase of the disk surface density at 0.55xRoche Limit, that is 1.6 Earth radii, close to our value of $R_L$. Below this location the disk is gravitationally stable as revealed by the value of the Toomre's Q coefficient displayed in their Figure 8.b. So, like in our simulation, the region below 1.6-1.7 Earth radii has a low effective viscosity. As mentioned in Takeda and Ida (2001), close to the planet, gravitational instability is suppressed because of the particles' incompressibility (they consider big bouncing particles, about 100 km across). The behavior reported in Takeda and Ida (2000) is very similar to what we observe in our simulations where the liquid incompressibility suppresses gravitational instability below $R_L$, inducing the formation of the dense inner disk.

The long term evolution of the protolunar disk is displayed in Figures 4, 5 and 6. The material remains mostly confined below $R_L$ (Figures 5, 6) and the disk cools down progressively. Thompson and Stevenson (1998) suggested that the disk may reach a steady-state in which the viscous heat production could perfectly balance the radiative losses through the photosphere. Here this delicate balance takes place only in the region $r>R_L$ where the material is in solid form, but this region contains less than 1% of the total disk mass. Below $R_L$, the disk is dominated by radiative cooling and viscous heating is negligible (Figure 4, column 3). As a consequence the temperature decreases efficiently with time in this region (Figure 4, Figure 5). Note however that the effective cooling timescale of ~$10^5$ years (Figure 6) observed in this simulation is somewhat longer than the typical radiative cooling timescale $T_c \sim (C_{vl} \ \Sigma \ T^4 / 2\sigma_S T_{irr}^4)$ that is about $2 \ 10^4$ years (using T~3000 K and $\Sigma \sim 10^9$ kg/m$^2$). This is because of a heating process appearing as series of transient hot and dense bursts in Figures 5 and 6, between $10^4$ and $10^5$ years that are induced by the liquid-solid phase transition. When the material in the hot and compact liquid disk cools down below Ts (1300K), it transforms into solid particles and form a granulate disk which can be gravitationally unstable even below $R_L$ (see section 2.6.1). Since a gravitationally unstable disk has a high effective viscosity, the viscous heat production increases suddenly and transient "heat bursts" are triggered as well as a short and intense phases of viscous spreading (visible on Figure 5 and 6). The time evolution of a typical heat burst is presented in details in Figure 7. The disk temperature is thus maintained at around 1300K on a timescale much larger than the standard cooling timescale because of the release of latent heat during the liquid-solid phase transition. The phase transition and associated viscosity and temperature increase might not be as sharp as modeled here since the liquid should crystallize over a temperature interval and not at a given temperature. If the existence of these bursts may be questioned, the temperature should be buffered close to the temperature of liquid-solid phase transition because of the associated change in viscosity, which should still increase the timescale for disk cooling.

The evolution of the disk's mass budget is displayed in Figure 8.a. The mass of the gravitationally unstable disk is the sum of the unstable liquid mass (beyond $R_L$) and the mass in solid form for which Q<1 as a function of time. In the first ~10 years the disk material (red line) flows inward and accumulates in the gravitationally stable region, between $R_\oplus$ and $R_L$. This is why the gravitationally unstable mass (black line) decreases. The total disk mass does not evolve much for $10^4$ years, while the mass in gravitationally unstable form (black line) decreases. After $10^4$ years spikes in the mass of gravitationally unstable material appear due to the heat bursts described above. During these bursts, a fraction of the mass flows from the gravitationally stable region (below $R_L$) to the gravitationally unstable region (beyond $R_L$) and also on Earth. At $10^5$ years, the disk is entirely solid and has a high viscosity, even below $R_L$, consequently the material flows onto the planet.

**3.3.2 A disk with 80 wt. % vapor**

The case of a disk starting with 80 wt. % vapor follows essentially the same steps (Supplementary Material 1) as most of the vapor condensates into solids in about 10 years, and since vapor is assumed to be non-viscous here, vapor is mostly static.

In both cases, there is never enough mass in the gravitationally unstable region to directly form the Moon. This could be an artifact due to the non-taking into account of a growing moon in our model. Indeed, a proto-moon at the disk edge should promote outward flow because of (1) its gravitational pull onto the disk in contact to its hill sphere and (2) the formation of a sharp edge at the Roche Limit may induce a strong viscous flow directed outward. This point is discussed below.

**3.3.3 First insight on the action of a growing proto-moon**

In order to have a first insight on the effect of a growing proto-moon on the disk's evolution, we have run a simulation of a disk finishing abruptly at the Roche Limit and starting with 20 wt. % vapor. This should mimic roughly the effect of disk truncation at the Roche Limit because of moon accretion (Kokubo and Ida 2000). We assume that all material crossing the Roche Limit is incorporated into a proto-moon. We could run such a simulation for about 10 years only, because of the reduced time-step imposed by the presence of the sharp edge. The mass evolution of the growing moon is displayed in Figure 9 and we see that it reaches about 30% of a moon mass in about 10 years. This mass is about 10 times larger than the mass of material in the unstable disk in the moon-free simulations (see above). So it seems that the presence of a sharp edge at the Roche Limit forces a strong outward spreading of the disk.

The material implanted in the proto-moon is initially in liquid form. Concerning the disk's evolution, as in the previous cases, a hot and compact liquid disk forms below 1.7 Earth radii (Figures 8 to 10 of the supplementary online material). A vapor rich disk develops between 2 and 3 Earth radii, due to the rapid removal of the disk's liquid component beyond 2 Earth radii that is rapidly transported to the proto-moon (Figure 8 and Figure 9 of the supplementary online material).

Thus, the effect of a growing proto-Moon on the disk evolution may be significant though the physics of accretion and tidal interaction of the proto-moon with the disk remains to be investigated. The next step will be to use the method of Charnoz et al. (2012) in order to compute the gravitational torque exerted by the proto-Moon onto the disk at every first order mean motion resonance.

**3.4 Evolution of a protolunar disk with a viscous vapor layer**

We now turn to the case of disks with a vapor layer that is abnormally viscous compared to the molecular viscosity. This could be made possible, as mentioned in section 2.7, by the onset of the magneto-rotational instability that is favored in ionized disks.

In Figure 10 and 11 is presented the evolution of a disk starting with 20 wt. % vapor and in which the vapor experiences MRI with the $\alpha$ turbulent parameter sets to $10^{-4}$ (see section 2.6 and Equation 18). The case of a disk starting with 80 wt. % vapor is presented in Supplementary Material 1. In rows 1 and 2 of Figure 10 the short term evolution of the disk is presented. Due to its high abnormal viscosity, the vapor layer produces a lot of heat. At the Earth's surface the liquid vaporizes due to the strong shear between the rotating Earth and the inner edge of the protolunar disk. This induces a strong inflow of material onto the Earth's surface (see the radial mass flux, Figure 10, column 4, rows 1 and 2) and vaporization. As a result the liquid disk "detaches" from the Earth's surface and is replaced by a vapor rich ring between 1 and 2 $R_e$. This effect is also clearly visible in the top of Figure 12. This vapor rich disk lasts about ~5-10 years. Then, the disk cools down and becomes liquid.

Because of its high viscosity the vapor efficiently flows down onto the Earth's surface (whereas it is not possible for the liquid layer since it is gravitationally stable below $R_L$) or escapes outward (Figure 7, row 4). As a consequence, a larger fraction of the disk mass falls back onto the Earth, and after 10 years the mass remaining in the disk is about 10 times less massive in the turbulent case than in the non-viscous case (compare the red lines in Figures 8.a and 8.b with Figures 8.c and 8.d). The cooling timescale is about 10 times shorter than in the non-viscous case: the heat bursts (that delay disk solidification) happen after about 100 to 1000 years evolution only. Disks starting with 20 wt. % vapor and 80 wt. % vapor have roughly the same long term evolution (Supplementary Material 1). The main difference between the two cases is the width of the vapor disk that forms above the Earth's surface. For the 20 wt. % vapor case the vapor disk extends up to about 2.5 Earth's radii (Figure 15) while it extends to about 3.5 Earth's radii for the 80 wt. % vapor case (Figure 18) after $10^3$ years.

**3.5 Kelvin Helmholtz instability and possible mixing of layers**

The work presented above assumes that at every moment, the two phases (vapor and liquid) are well separated, which may be a matter of debate. If the two phases are well mixed, the disk sound velocity may drop significantly, making the disk prone to gravitational instability (Kieffer 1977, Thompson & Stevenson 1988, Ward 2012). Substantial mixing may occur between the two phases, in particular because of the onset of the Kelvin Helmholtz instability. Kelvin Helmholtz (KH) instability can occur because of the difference in the orbital velocity between the liquid layer and the vapor layer at their interface. What happens in the KH turbulent layer of the fluid is however very unclear. On the one hand, Machida & Abe (2004) assume that the KH turbulence prevents the onset of gravitational instability, like often assumed in models of planetesimal formation in protoplanetary disks (indeed, turbulence induced by KH instability may stir solid material and increase random velocities, which acts against gravitational instability e.g. Sekiya 1998 or Chiang 2008). On the other hand, one may argue that in the KH unstable layer, the vapor will be well mixed with the liquid (like petrol in car engines), creating a two-component medium with vapor at equilibrium with liquid in which the sound velocity *drops* very sharply (see for example Figure 2 of Ward 2012, or see extensive comments on this effect in Thompson & Stevenson 1988) and where gravitational instability could be possible. It is the very the opposite assumption of Machida & Abe (2004)! This problem would deserve a full dedicated study in itself, well beyond the scope of the present paper. In the current context we think it is very possible that the Machida & Abe (2004) reasoning, inspired from dust sedimentation in gaseous protoplanetary disk maybe wrong. Indeed dust in protoplanetary disk *are not made of the same material as the surrounding gas* so that dust grains are not in thermodynamical equilibrium with the surrounding disk's gas. In our context, the situation is qualitatively different since the droplets and vapor are two phases of the same material. If thermodynamical equilibrium is achieved, then the two phases medium should obey the Clapeyron relation and the sound velocity of the mixture may drop significantly (Thompson & Stevenson 1988).

In order to simulate the effect of the KH instability the following procedure is used: first, to compute the mass of the layer experiencing KH instability we adopt the same prescription as in Machida & Abe (2004) based on Sekiya (1998). The surface density of the turbulent layer is computed according to Equation 29 of Machida & Abe (2004) using a constant Richardson number $J_c=0.25$ and a keplerian default $\eta=(C_{sv}/r\omega_k)^2$. At every location, the liquid surface density can be splitted into a KH turbulent layer and a KH non turbulent layer so that $\sigma_l=\sigma_{l,KH}+\sigma_{l,NKH}$. The non KH turbulent part (NKH subscript) behaves like an incompressible fluid as described in section 2.6.2. So in Equation 17 $\sigma_l$ is replaced by $\sigma_{l,NKH}$. We assume that the turbulent KH layer has a two-phase sound velocity $C_{s2}$ (computed according to Equation 15 of Ward 2012). Since the vapor mass fraction, X, in the KH unstable layer is unknown we assume $X=10^{-3}$ when computing $C_{s2}$ (leading to low values of the sound velocity in case of mixing, see Figure 2 of Ward 2012). We then compute the Toomre parameter of the KH unstable layer according to $Q=\omega_k^2 C_{s2}^2/(G\,\sigma_{l,T})$. If Q > 1 then the KH unstable layer is gravitationally stable and its effective viscosity is $\nu_{KH}=0.001$ m$^2$/s (~molecular viscosity of a magma), if Q<1 then the KH unstable layer is gravitationally unstable and its effective viscosity is $\nu_{KH}=\pi^2 G^2 \sigma_{l,T}^2/\omega_k^3$ (the effective viscosity of a gravitationally unstable disk). And finally the "grand total" effective viscosity of the condensed layer, $\nu_l$, is computed as the average viscosity weighted by the surface densities of the KH stable and KH unstable layers of the liquid phase, i.e $\nu_l=(\sigma_{l,NKH}\nu_{KH}+\sigma_{l,KH}\nu_{KH})/\sigma_l$.

In order to illustrate the possible effect of KH instability we present the case of a disk starting initially with 20 wt. % vapor mass fraction and a non-viscous vapor phase (Figures 12 to 15), to compare with the case discussed in section 3.3). In Figure 13 the local mass fraction of the gravitationally unstable material is plotted as a function of distance at two different epochs. In the region below $R_L$ (1.7 $R_e$) we observe that only 10% of the disk mass is locally unstable (the KH instability does not penetrate down to the disk midplane because of the very high density at the midplane). Due to its low sound velocity, the mixture becomes gravitationally unstable close to the Earth and has a high effective viscosity. As a result and because of the strong shear between the disk with the Earth's surface, the material is turned into vapor. So the effect of the KH instability is mainly localized close to the Earth's surface, whereas it does not affect the totality of the disk mass.

Inspection of the evolution of the disk surface density (Figure 16. top) and mass vapor fraction (Figure 16. bottom) shows that the usual inward flow of liquid accumulating between the Earth's surface and $R_L$ is found. Because of the onset of KH turbulence, the liquid is more prone to gravitational instability, and thus, more viscous. As a result, the liquid/gas mixture below $R_L$ heats up and the liquid evaporates almost completely (70% vapor mass fraction) just above the earth's surface. So a vapor rich ring appears just above the Earth and lasts about 10 years. After 10 years, like in the standard case, most of the vapor has disappeared and the disk mass is mainly contained in the liquid phase. On longer timescales, the KH allows a fraction of the disk to be gravitationally unstable especially close to the Earth's surface. Because of the shear with the Earth, vapor is generated close to the Earth's surface that mixes with the liquid and makes it gravitationally unstable. As a result, the innermost part of the disk (below 1.2-1.3 $R_e$) flows down to the Earth's surface after 1000 years. So we end up with a liquid ring, detached from Earth, that extends from 1.3 to 1.7 $R_e$. We find that more mass ends up on the Earth's surface in the KH case than in the standard case (compare with H and Figure 14). After $10^4$ years evolution, solidification occurs and the same bursts are found as in the standard cases. So not surprisingly, we find that when the two layers are allowed to mix, the innermost portion of the disk falls more rapidly onto the Earth but the disk general evolution is not fundamentally changed. The total mass of material in the gravitationally unstable layer is never large enough to make the Moon with its present mass if the growing moon is not accounted for.

## 4. Evolution of volatile and refractory species

One of the most remarkable properties of the lunar material is its scarcity in volatile elements (see the recent review by Taylor and Wieczoreck 2014) and in some light isotopes like Zn (Paniello et al., 2012). The D/H ratio of the Moon's water is higher that the Earth's by up to a few hundred per mil (Greenwood et al., 2011). Whereas it has been proposed that such a depletion could be due to hydrodynamical escape of light species in the protolunar disk (Machida & Abe 2004, Desch & Taylor 2012) the relatively high abundance of heavy oxygen atoms and short lifetime of the disk may prevent the escape of the lightest species in a diffusion-limited hydrodynamical-escape model suggesting that the Earth-Moon system is essentially closed during the Moon formation process (Nakajima et al., 2014b). So it is interesting to quantify if the mechanisms explored in the present paper may, or may not, lead to substantial modifications of the light to heavy elements mass ratio. Fractionation may be expected in a stratified disk: since the viscosity of the liquid or solid phase (that bears heavy elements) is always different from the viscosity of the vapor (that bears light elements), the two phases have different dynamics. In consequence a dynamical modification of the light/heavy elements mass ratio is expected. Ideally we would like to use a real thermochemical model for the computation of the abundances of the major elements (like, Si, O, Na, Zn etc.) like in Visscher & Fegley (2013) and track the evolution of their abundance in the dynamical model described above. Unfortunately, the high temperature and pressure ranges met in the current problem extend well beyond the validity of the thermodynamical tables (>4500K). So in order to get an idea of the evolution of the system's content in volatile and refractory elements, we propose a toy model that distinguishes between different degrees of volatility.

We assume that the disk composition is always dominated by Si and that it contains only some traces of other species characterized by different degrees of volatility K. For a given hypothetic specie with volatility K ("specie **k**" for short), let $C_v(K)$ the local mass concentration of specie **k** in the vapor layer and $C_l(K)$ the local mass concentration of specie **k** in the liquid layer. K is defined as:

$$K = \frac{C_l(K)}{C_v(K)}$$  **Equation 20**

So K is simply the ratio of the surface density of specie **k** in the liquid phase by its surface density in the vapor phase. We assume that for a given specie, K is independent of temperature and pressure (which may be unphysical, but it is a zero order model). K may be seen as a volatility coefficient: for K< 1 the specie is volatile and for K>1 the specie is refractory. Because K is supposed to be constant it is not straightforward to link K to a real physical specie. However, it is natural to expect that very refractory species, like thorium, should behave like K>>1, moderately volatile species like potassium may have K<1, and very volatile elements like Zinc may have K<< 1.

We have tracked 7 species with different values of K: $10^3$, $10^2$, 10, 1, 0.1, 0.01, 0.001. For each specie their surface density in the vapor, $\sigma_v(K, r)$, and in the condensed phase, $\sigma_l(K,r)$ are tracked for each location r. Their surface density evolution is computed as follows

1) At the beginning a time step a specie is advected with the local flow velocity as computed in Equation 2. $\sigma_v(K, r)$ and $\sigma_l(K,r)$ are modified according to the mass conservation equations (eq.1).

2) At the end of each time step, $\sigma_v(k,r)$ and $\sigma_l(k,r)$ are modifed for every location r so that their ratio K is respected (as in eq.20) while ensuring mass conservation.

Using this simple method we can track the spatial abundance of species. Finally to summarize our results we compute the total mass of each element in the disk. Whereas cases with a viscous vapor layer (Figure 18) and a non-viscous vapor layer (Figure 17) show very different evolutions, the influence of the initial amount of vapor is moderate.

When the vapor is non-viscous, it is dynamically almost static (compared to the condensed phase). As the liquid cannot flow onto the planet because it is stable below $R_L$, the system is closed so the ratio of the different species to the most refractory one is constant for about $10^4$ years (Figure 17). After a few $10^4$ years the ratio of volatile/refractory elements increases moderately, and we end up with a *volatile enriched* disk! This is a direct consequence of the following processes: (1) the vapor is almost static due to its low viscosity (2) as the disk cools down more and more material goes into the condensed phase. After $10^4$ years, the liquid transforms into solid (after a few heating bursts driven by gravitational instability, see section 3.3.1), that can become gravitationally unstable even below $R_L$ and flow onto the Earth. As a result the refractory material flows toward the proto-Earth, leaving behind a disk slightly enriched in volatile elements and trapped in the remaining vapor close to the Earth's surface. This explains simply why the volatile/refractory elements mass ratio increases when the system solidifies.

We turn now to the case where the vapor is turbulent (Figure 18) which shows a radically different behavior. After about 1 year the disks shows a strong depletion of volatile elements lasting about 1000 years. Indeed, due to its high viscosity the vapor phase, and the volatile elements it bears, flows efficiently inward while the liquid phase (and the refractory elements it bears) cannot flow below $R_L$ because the liquid is stable below $R_L$. As a result a process resembling a distillation occurs in the disk: as the vapor layer is evacuated toward the Earth, because of the heat generated by its anomalous viscosity, the liquid phase feeds the vapor phase at the same time. As a result the remaining disk is strongly depleted in volatiles. When the disk solidifies, after $10^3$ to $10^4$ years due to its high viscosity, the solid phase eventually flows onto the Earth and the volatile fraction increases again, like in the non-turbulent case. The volatile abundance of the most volatile species (with K=$10^3$ here) are depleted by a factor between 100 and 1000 compared to the most refractory species (K=$10^{-3}$), in qualitative agreement with the range of volatile depletion measured in Lunar material (see e.g. Taylor and Wieczoreck 2014).

## 5. Summary and discussion

In the present paper we have detailed the evolution of the protolunar disk, coupling a dynamical and a thermodynamical model, over up to $10^5$ years, assuming that the disk is stratified with a condensed layer in the midplane, either liquid or solid, topped by a vapor layer. The gravitational instability is modeled as an artificial increase in viscosity that triggers an efficient transport of angular momentum and disk heating. The three phases (solid, liquid, gas) have very different viscosities (see section 2.5). Our hypotheses are:

- The vapor phase may be either non viscous (i.e low molecular viscosity) or abnormally viscous (like in the case of magneto-rotational turbulence) and the two cases are investigated using an $\alpha$ model.
- The liquid phase, assumed to be incompressible, is never gravitationally unstable below a critical distance, $R_L$~1.67 Earth's radii following Sekiya (1983), so the liquid viscosity is always very low below $R_L$ and very high beyond $R_L$ (because of gravitational instability).
- The solid phase may be gravitationally unstable (i.e high viscosity) or not (i.e. low viscosity) at any distance from the Earth, depending on the local value of the Toomre coefficient Q.

We model the coupled evolution of the two phases (vapor and condensed) assuming vertical hydrostatic equilibrium and thermodynamical equilibrium. We do not model Moon's formation in our standard simulations. In total five cases were studied:

- A disks starting with 20 wt. % vapor that is non viscous
- A disks starting with 20 wt. % vapor that is viscous with $\alpha=10^{-4}$ (due to MRI)
- A disks starting with 20 wt. % vapor that is non viscous, but assuming that Kelvin Helmholtz instability mixes substantially the liquid and the gas layer resulting in a sound velocity drop.
- A disks starting with 80 wt. % vapor that is non viscous (Supplementary Material 1)
- A disk starting with 80 wt. % vapor that is viscous with $\alpha=10^{-4}$ (Supplementary Material 1)

We provide one additional simulation with an ersatz of moon-accretion by imposing the presence of a sharp outer edge at the Roche Limit and assuming that all material crossing the Roche Limit is incorporated in a proto-moon (Figure 9, Supplementary Material 3).

### 5.1 General behavior of the disk: formation of a hot compact disk below 1.7 Earth radii

In all cases, the protolunar disk evolves through 3 distinct stages:

- Stage 1: The disk cools from the exterior to the interior. In about 10 years the vapor layer condenses into liquid. Most of the disk material flows inward and accumulates in a compact disk extending from the Earth's surface to $R_L$ (~1.7 Earth Radii) i.e. the frontier of the gravitational instability zone for an incompressible fluid disk according to Sekiya (1983). This hot compact liquid disk is gravitationally stable and has a low effective viscosity. It remains in orbit for an extended period of time; still, a small fraction of material falls on Earth.
- Stage 2: The hot and compact liquid disk reaches a steady state during $10^4$ to $10^5$ years (for viscous and non viscous vapor respectively).
- Stage 3: the disk solidifies when its temperature drops below 1300K. Several short and transient heat bursts are triggered that heat up the disk and substantially lengthen its cooling time, due to the sudden increase in viscosity when the disk condenses into solid particles and the sudden release of latent heat. During these heat bursts, the material flows efficiently down to the Earth's surface and beyond $R_L$. After these bursts, the disk ends in solid form.

Inspections at the heat production rate and radiative cooling show that a power-balance is never achieved: the radiative cooling always dominates the viscous heating contrary to what is often assumed (Thompson & Stevenson 1988, Ward 2012, Canup & Salmon 2012) and as recently noted by Ward (2014a, b).

### 5.1 Effect of an abnormal viscosity (MRI) for the vapor phase

This general picture must be somewhat modified if the vapor is viscous. Series of simulations assuming that the vapor layer is viscous using $\alpha=10^{-4}$ show that:

- Due to the vapor's strong viscosity, the disk is volatilized almost fully in the first 1-10 years of evolution and the vapor mostly flows inward.
- A hot vapor atmosphere, devoid of liquid material, settles at the junction of the disk inner edge due to the strong difference in velocity between the Earth and the disk
- More mass falls onto the Earth than in the inviscid vapor case. As a result the hot and compact liquid disk remaining in orbit is about 10 times less massive than in the inviscid vapor case.
- Due to its reduced mass, the hot and compact liquid disk cools down to the solidification temperature in a shorter time (about 1000 years) when the bursts are triggered.
- The disk is finally completely solid after ~$10^3$ years (rather than $10^5$ years in the inviscid vapor case).

**5.2 Effect of mixing of the two layers due to Kelvin Helmholtz instability**

Whereas Machida & Abe (2004) found that the sedimentation timescale of droplets is very short, about $10^{-2}$ years, favoring a stratified disk, Kelvin Helmholtz instability at the interface of the vapor and liquid layer may lead to substantial mixing of the gas and liquid. The mixing of a small amount of gas into liquid results in a strong decrease of the mixture sound velocity (a classical effect in a vapor/liquid mixture, see e.g. Kieffer et al., 1977) and thus favors the onset of gravitational instability as depicted in Thompson & Stevenson (1988) or Ward (2012). Using the Machida & Abe (2004) formalism to compute the mass of the Kelvin Helmholtz unstable layer we find that, at most, 10 wt. % of the hot compact liquid disk (below 1.7 $R_e$) can become unstable (Figure 11). The instability is mostly located close to the Earth's surface because the shear here is strong and favors vapor production. The disk vaporizes close to the Earth's surface and up to 1.7 $R_e$. An intense flow onto Earth occurs, but does not change the global picture of the disk evolution.

**5.3 Depletion in volatile elements in disks with abnormally viscous vapor.**

Because of their different viscosities, the volatile (i.e. gas) and refractory (i.e. condensed) phases have a different dynamics which can lead to an efficient dynamical fractionation process in between refractory and volatile elements.

- If the vapor layer is not viscous, it remains quite static and the final disk appears slightly enriched in *volatiles* after disk solidification due to the fall of solid material (rich in refractory elements) onto the Earth's surface (Figure 17.top) and the survival of a vapor layer in orbit.
- In the case where the vapor layer is viscous, and thus mobile, most of the volatile elements fall onto the Earth's surface in the first 10 years, and the resulting orbiting material is enriched in *refractory* elements by a factor of 10 to 100, like for the today's Moon. When the disk solidifies the refractory elements fall on the Earth and the abundance of volatile elements rise again almost to their initial value, after $10^3$ to $10^4$ years.

Our major finding is thus that only an abnormally viscous vapor disk is able to deplete the disk in volatile elements by implanting them on the Earth. The disk is devolatilized when the condensed phase is still liquid, i.e before $10^3$ years, because only a liquid disk is able to remain in orbit and retain the refractory elements due to its low viscosity below $R_L$. From these considerations we see that for the Moon to be devolatilized it must form from the disk material between 10 and $10^{3-4}$ years when the disk is mostly liquid, and before its solidification. This is in agreement with the occurrence of an early magma ocean phase on the Moon as attested by many geological evidences, among them the presence of an old arnothositic crust at the Moon's surface. As a direct consequence, the present work suggests that the vapor phase was turbulent, which may imply the presence of an early magnetic field during the Earth's formation, coming either from the Sun or from an early terrestrial dynamo (Ziegler and Stegman, 2013) in order for the magneto-rotational instability to take place.

Note however that the conditions for MRI to be active are quite restrictive. Even under the presence of a magnetic field and with a disk material that is ionized, the presence of droplets may act as a source of electrons that could de-ionize the material and suppress MRI. However, if MRI is active, then it provides an efficient way to separate the volatile from the refractory material.

One of the outstanding questions about Moon's formation is the characteristics of the impactor that has major consequences concerning the lunar material composition. Pahlevan and Stevenson (2007) suggest that turbulent mixing in the disk, before Moon's accretion, may equilibrate isotopic species between the Earth and the disk. In order for this mechanism to be effective, they find that the disk should last about 1000 years assuming $\alpha=3 \cdot 10^{-4}$ and T=2500K (Pahlevan & Stevenson 2007). In the present context we find much higher temperatures in the first 10 years (about 5000 to 10000 K) so that the mixing timescale should reduce accordingly (since the diffusion coefficient D scales with T, assuming $D \sim \alpha \, C_s^2/\Omega_K$). So even in the case of a low vapor content, isotopic species may have time to equilibrate before Moon's assembling.

## 6. Conclusion: looking for the right conditions to make the Moon

In most cases considered above, we find that that most of the disk falls back onto the Earth rather than being transported beyond the Roche Limit, that is the natural formation place of the proto-moon. However, effects not considered here may change the dynamics of the system:

- A growing proto-Moon orbiting exterior to the disk may attract a significant mass of material outside the Roche Limit. A first test of this scenario is presented in section 3.4 and in the Suplementary Material (section 3) where the presence of a proto-moon is simulated through an artificial sharp edge at the at the Roche Limit. We find that 10 times more mass crosses the Roche Limit than in the moon-free case. We observe the formation of a 1/3 moon mass object in 10 years simulation. Additionally, we find a high vapor fraction the disk's edge because of the rapid transport of liquid beyond the Roche Limit. This indicates that the presence of a proto-moon significantly affects the disk evolution, and will be studied in a further study.

- In the present work an Earth's rotation period of 4h is assumed like the in Canup (2004) standard model. Since the orbital period at the disk inner edge is about 2h only, this results in an intense shear that promotes the fall of material onto Earth. However, in some recent models (Cùk & Stewart 2012), a rapidly rotating Earth is assumed (about 2h, close to the centrifugal instability). This should lower the shear and favor the preservation of material into the disk and facilitates the global circulation of material in the Earth/Disk/Proto-Moon regions more easy.

- A simple alternative is that, just after the impact, most of the disk mass is projected *beyond* the Roche Limit, rather than *inside,* as hypothesized in the current paper or in Kokubo et al. 2000, or in Salmon and Canup (2013). In that case, the proto-Moon would form very rapidly (in a few orbital periods only) as the condensed phase is always gravitationally unstable beyond the Roche Limit. Whereas this seem quite a straightforward solution it is not clear for the moment if this hypothesis is true as different simulations show different mass repartitions after the impact (compare for example Canup 2004, 2012 or Nakajima & Stevenson 2014a). However even if this mechanism is the right one, this would leave completely open the question of the volatile depletion of the Moon because, on such a short timescale, an efficient volatile depletion is very difficult (Nakajima and Stevenson 2014b) .

- A last, and very interesting, alternative would be that the Moon was not formed in a single impact, but rather that it was progressively assembled after several impacts as suggested in Citron et al (2014). As moons always form after an impact, and as smaller disk/planet mass ratio promotes the formation of multiple moons systems (see Crida & Charnoz 2012) we may imagine that a retinue of moonlets was formed after multiple moderate impacts and that they progressively gathered into a single object as the Earth's tides force an expansion of the moonlets orbit. Such a scenario is still to be investigated in details. Note that due to the Earth's fast rotation and strong flattening after each impact, the debris disks should settle in the Earth equatorial plane due to strong Earth's J2.

Generally speaking due to the intrinsic simplicity of the physical model used here, this work must be taken with care and must be seen as a first attempt to track quantitatively and numerically the coupled evolution of the protolunar disk dynamical and thermodynamical evolution. Many uncertainties on the disk physics remain, in particular concerning the mixing of droplets with gas, which prevent any definitive conclusions. There are also some uncertainties about the effective viscosity of the liquid and the present work has adopted the viscosity prescription of Machida and Abe (2004) of a perfect non-compressible fluid, resulting in a stable disk below 1.7 Earth's radii. As mentioned earlier, some authors propose that surface tension favors big droplets suggesting that a behavior closer to a particle disk would be possible. In that case we would expect that the liquid disk would be gravitationally unstable all the way down to the Earth's surface, so that an intense inward flow may take place, promoting accretion onto Earth. This would also shorten considerably the disk's lifetime.

The size of particles in the solid phase may also potentially play a role. We have arbitrarily assumed here that grains are 1m in radius. Increasing the particle size should not change the disk's physics in the gravitationally unstable part of the disk (because in the GI part, everything is controlled by the surface gravity only, see Tanaka and Ida 2001). Conversely, close to the planet, the disk dynamics may change as the viscosity is due to collisions with particles (see section 2.5.2) and it scales with the optical depth which is proportional to $1/r_g$. So increasing the particle size may decrease the viscosity close to the planet and decrease the mass flow onto the Earth. Conversely, if the proto-earth had an atmosphere, gas-drag from the atmosphere may lead to a substantial increase of the mass flux on the Earth-surface.

In conclusion, our present work shows that the thermal and dynamical evolution of the protolunar disk may be quite complex and several supplementary effects must be studied. We hope this work helps to raise somewhat the veil on the protolunar disk structure and evolution that is one of the most mysterious structures of planet formation. Possible extension of this work includes the study of circumplanetary disks around ice giants (Uranus and Neptune) as their satellite systems may be formed in the aftermath of giant impacts, at the end of planet formation.

**ACKNOWLEDGEMENTS :**

We thank two anonymous reviewers for their useful comments. We acknowledge the financial support of the UnivEarthS Labex program at Sorbonne Paris Cité (ANR-10-LABX-0023 and ANR-11-IDEX-0005-02). This work was supported by Université Paris Diderot and by a Campus Spatial grant. Sebastien Charnoz thanks the IUF (Institut Universitaire de France) for financial support.

# FIGURES

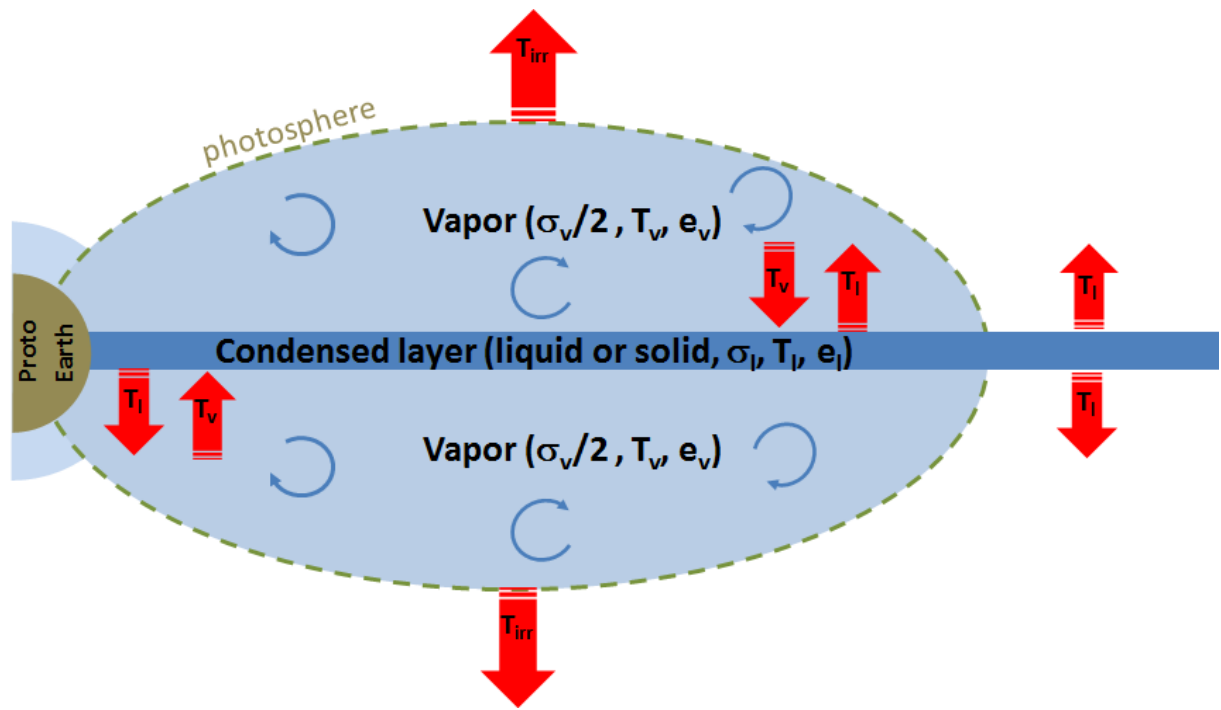

**Figure 1**

Sketch of the system considered in this paper: a condensed layer (liquid or solid ) layer topped by two vapor layers. Red arrows indicate the directions of radiative energy transport and the temperature of emission. The three layers exchange energy at their own current temperature and ultimately the energy is radiated to space by the vapor's photosphere with an effective temperature $T_{irr}$= 2000K. When the condensed disk is too cold for a vapor layer to be present it radiates its energy to space with its effective temperature ($T_l$).

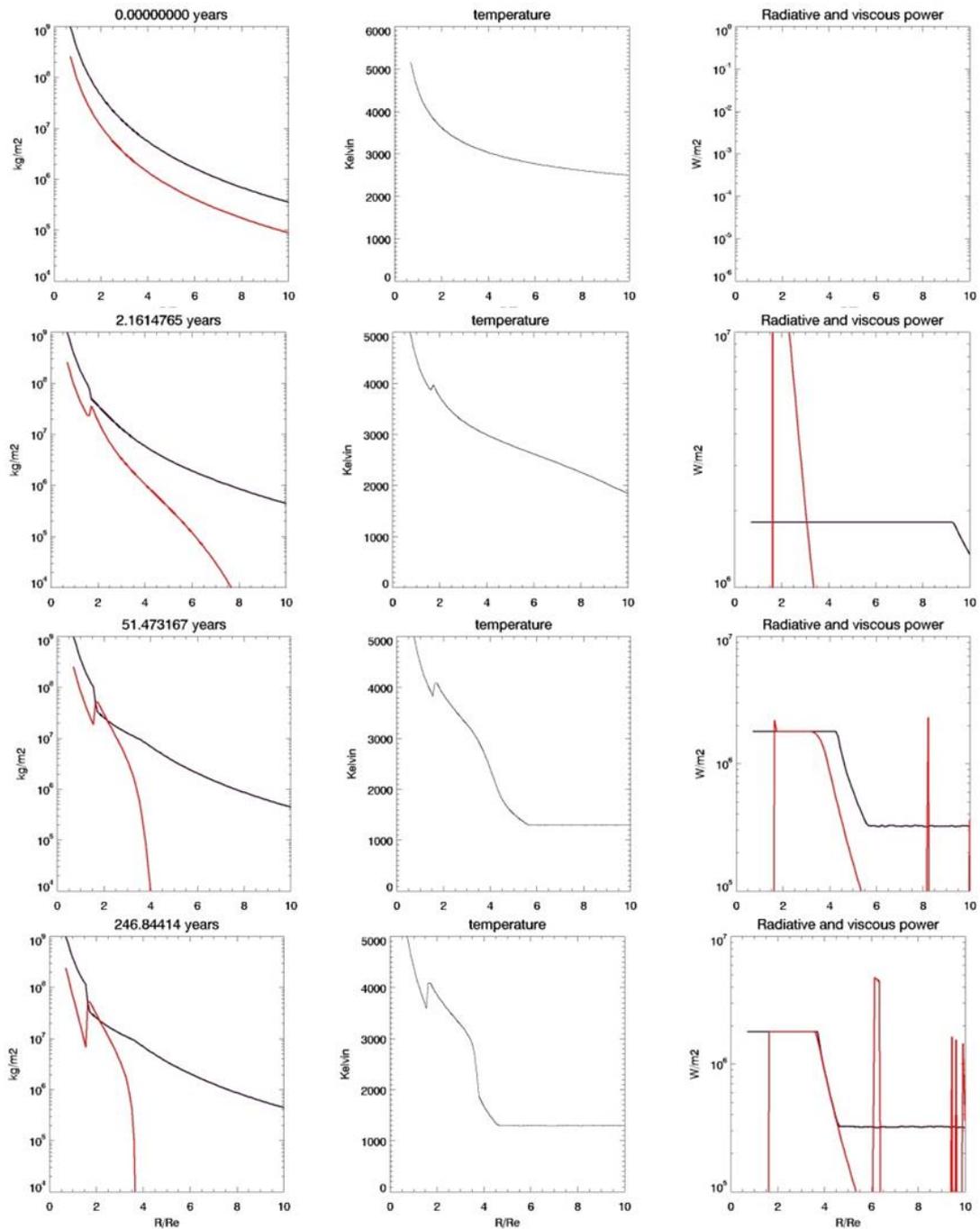

**Figure 2**

Time evolution of a protolunar disk neglecting its dynamical evolution and assuming 20 wt. % initial vapor fraction. Each row shows the system at different epochs. The first column shows the surface density as a function of time (the black line is for the condensed phase surface density, while the red line stands for the vapor phase), the second column shows the temperature, and the third column shows the power emitted through radiations at the disk surface (black line) and the power generated through viscous dissipation (red line). Spikes in the viscous power emissions are due to local transitions from liquid to solid that strongly increases the local viscosity.

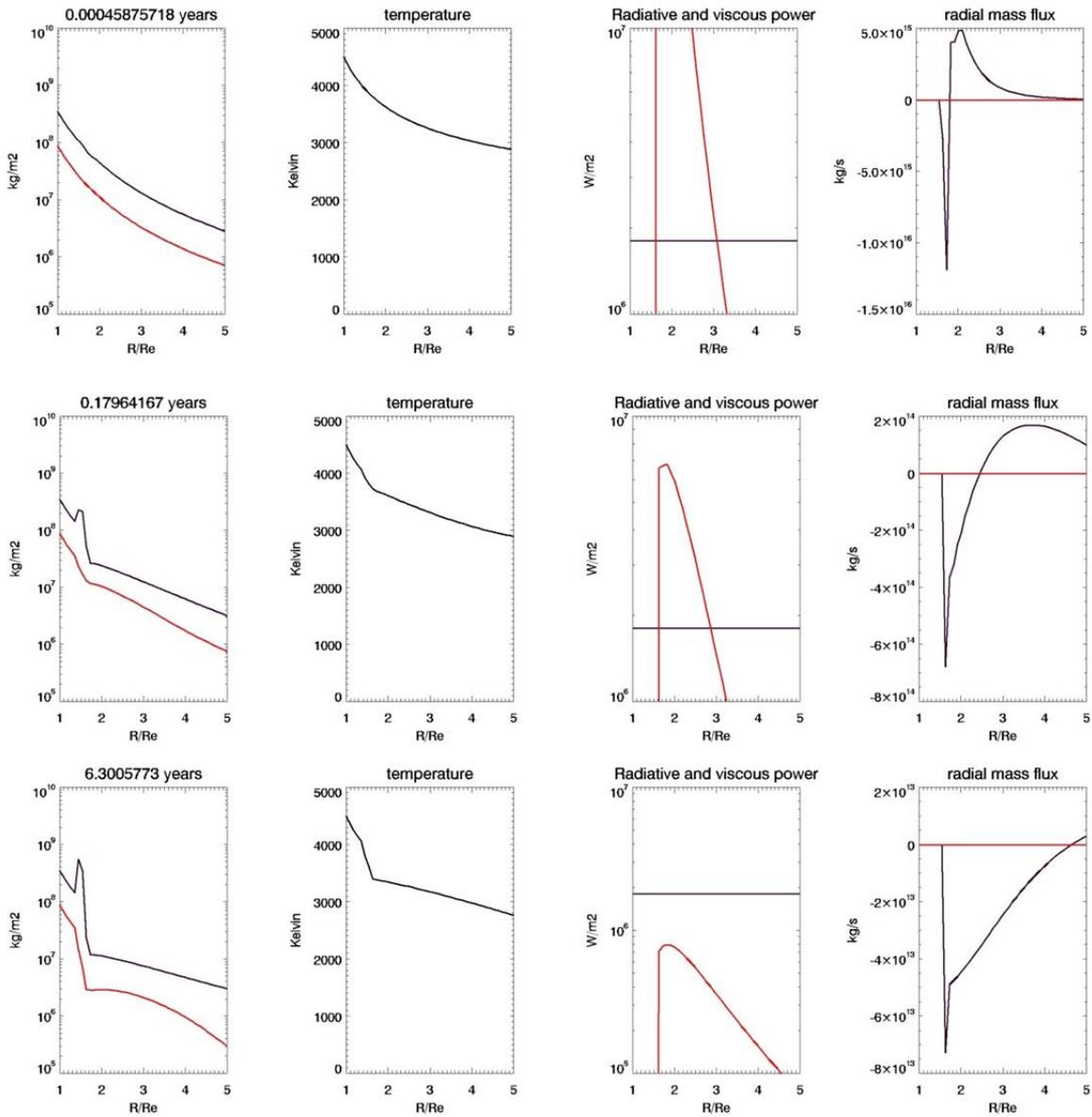

**Figure 3**

Short term evolution of a protolunar disk with 20 wt. % vapor initially. Rows (a) to (c) correspond to different dates. Column 1: surface density (black line: condensed phase, red line: vapor phase); column 2: midplane temperature; column 3: radiative emission from the disk (black line) and viscous power heating (red line); column 4: radial mass flux (black line: condensed phase, red line: vapor phase).

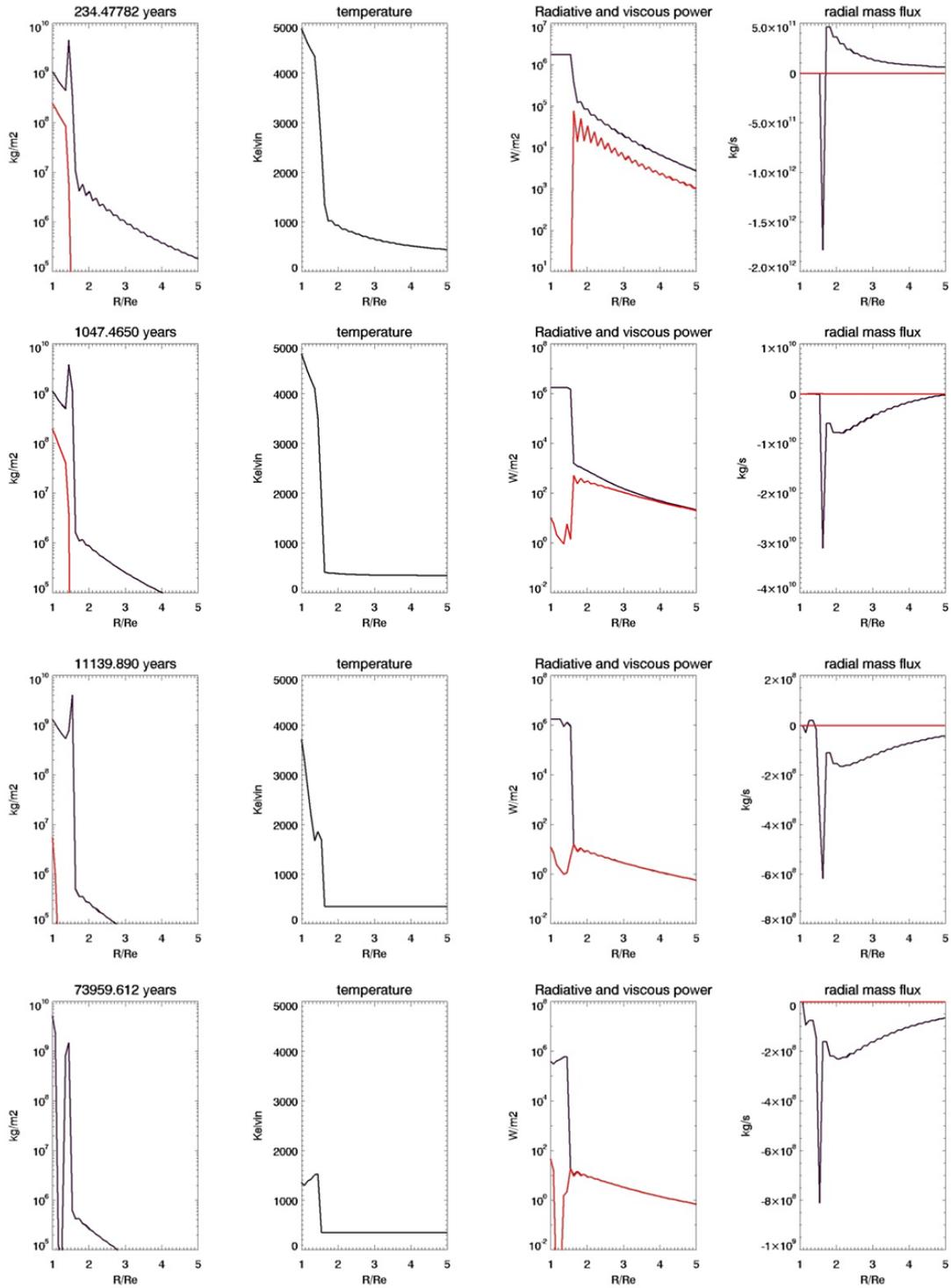

**Figure 4**

Long term evolution of a disk with 20 wt. % vapor initially and a non-turbulent gas phase. See figure 3 for legend.

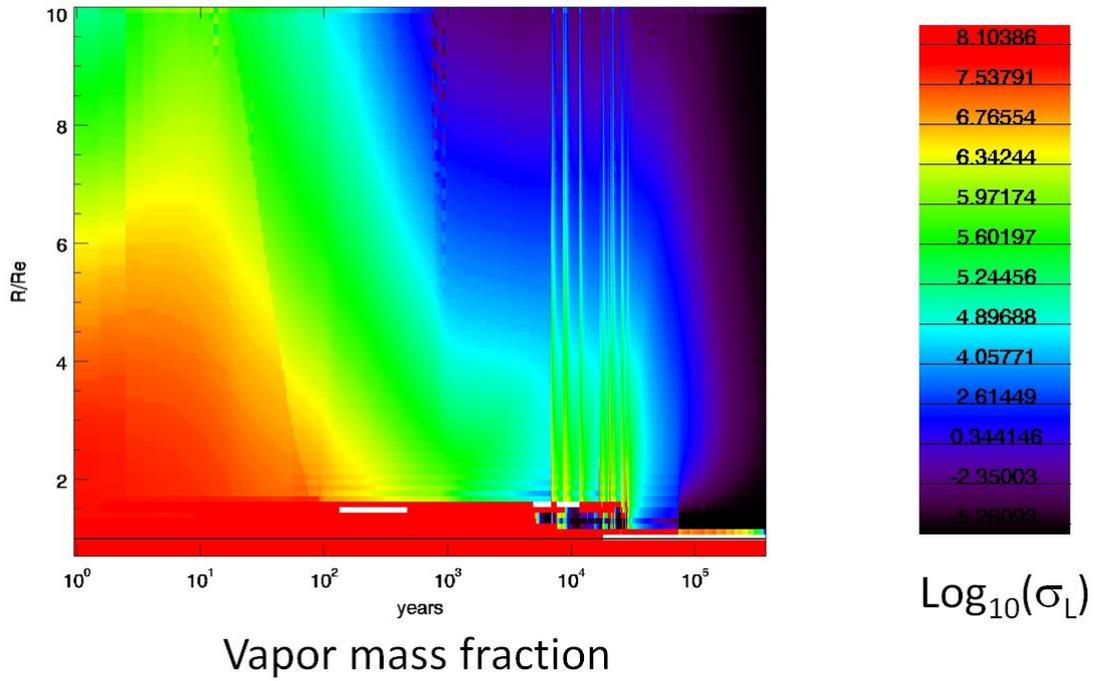

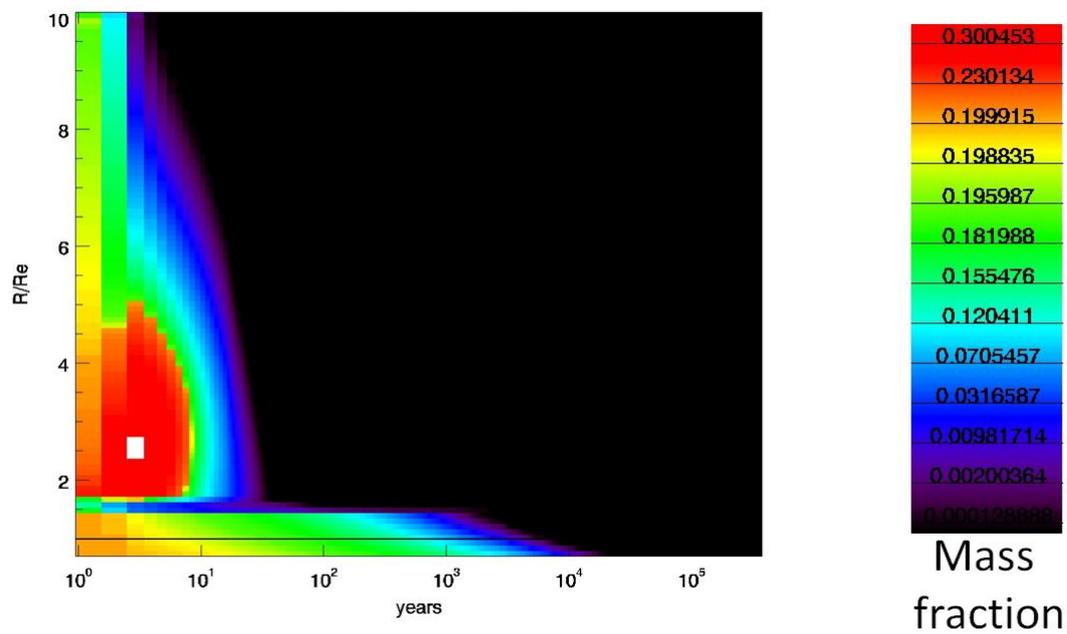

**Figure 5**

Time evolution of a disk with 20 wt. % vapor mass fraction initially and a non-viscous vapor. Top: surface density of the condensed phase; bottom: vapor mass fraction.

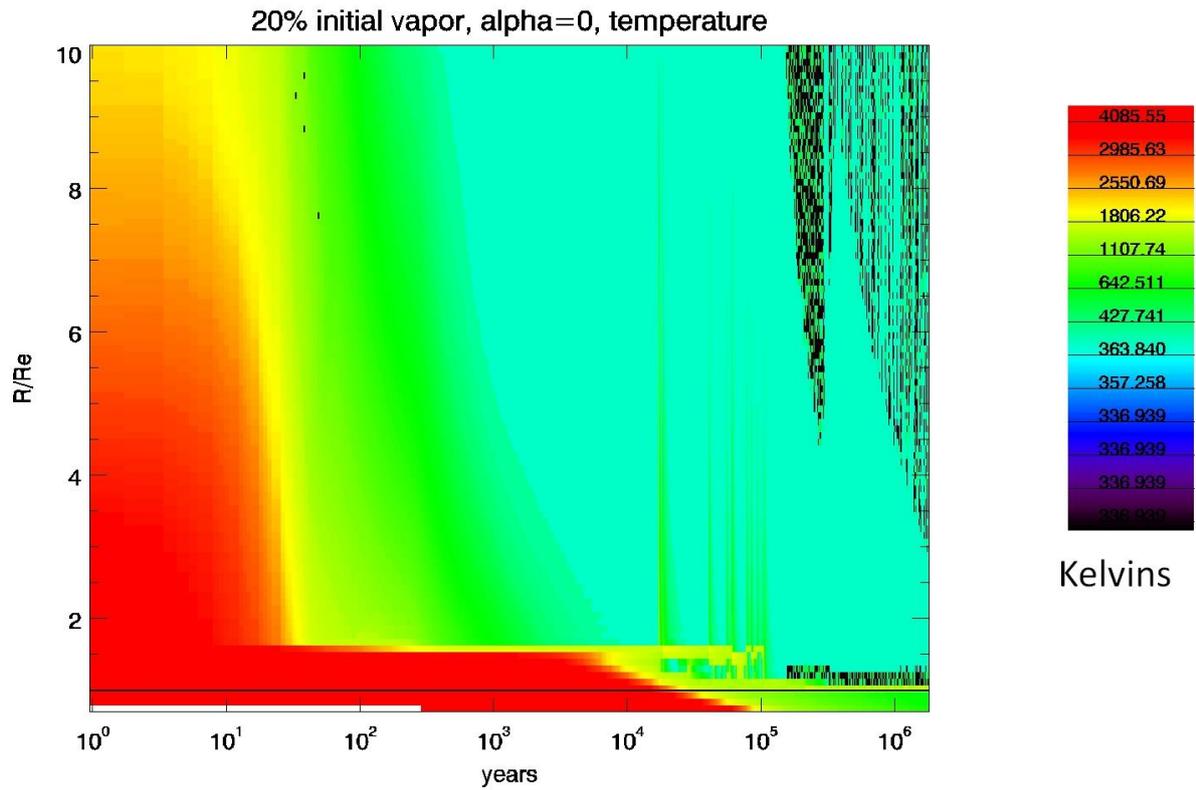

**Figure 6**

Temperature evolution of the protolunar disk in the case of 20 wt. % initial vapor fraction and a non-viscous vapor ($\alpha=0$). Heat bursts are clearly visible between $10^4$ and $10^5$ years. Note that the majority of the disk mass is contained between $R_\oplus$ and $1.6 R_e$ after 10 years evolution and is gravitationally stable when the disk is fluid. This is why this region cools much more slowly that the rest of the disk.

# Gravitational instability driven thermal burst

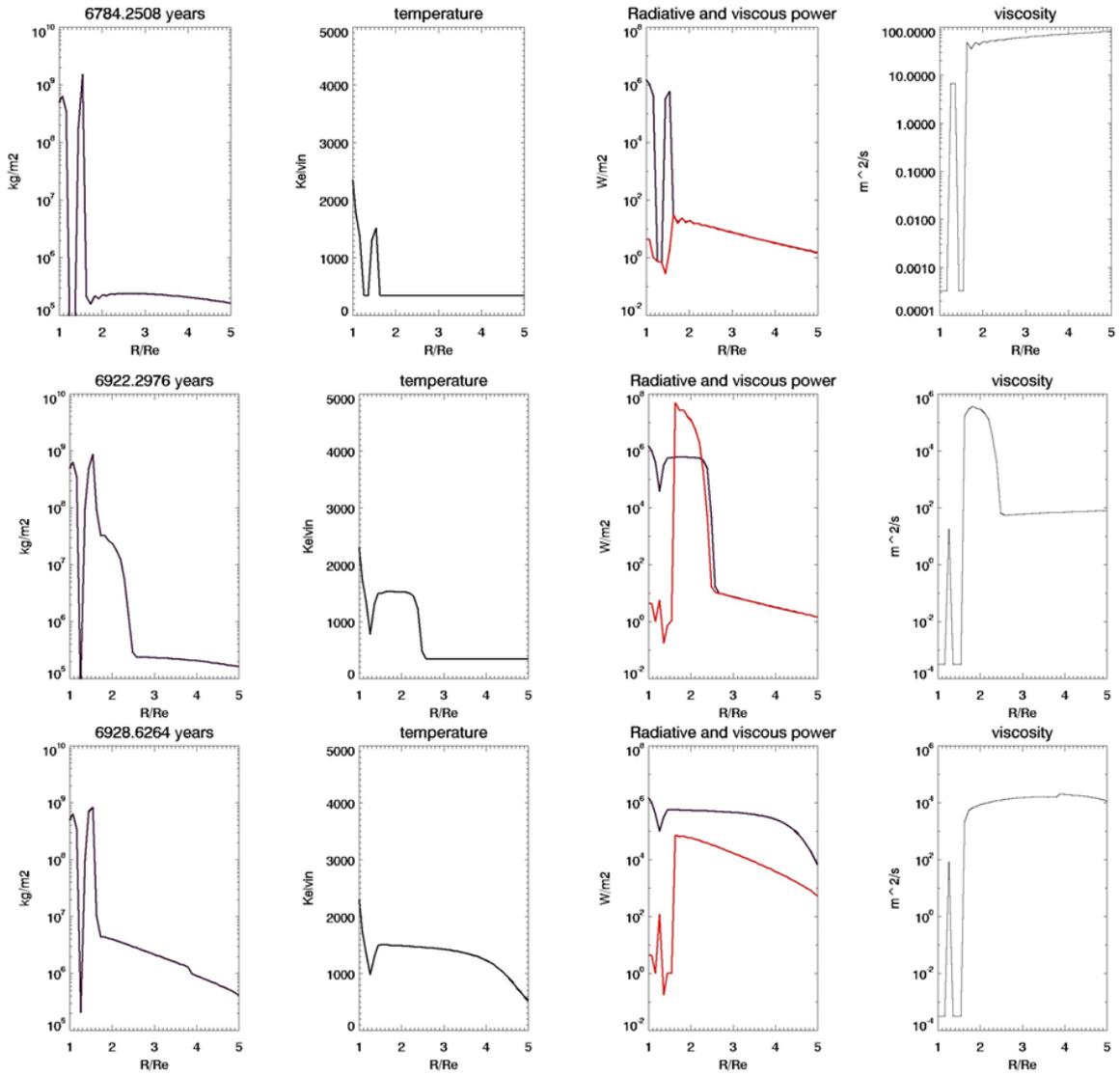

**Figure 7**

Time evolution of the disk during a thermal burst. Columns 1 to 3 are the same as in Fig. 5. Column 4 shows the local viscosity of the condensed phase.

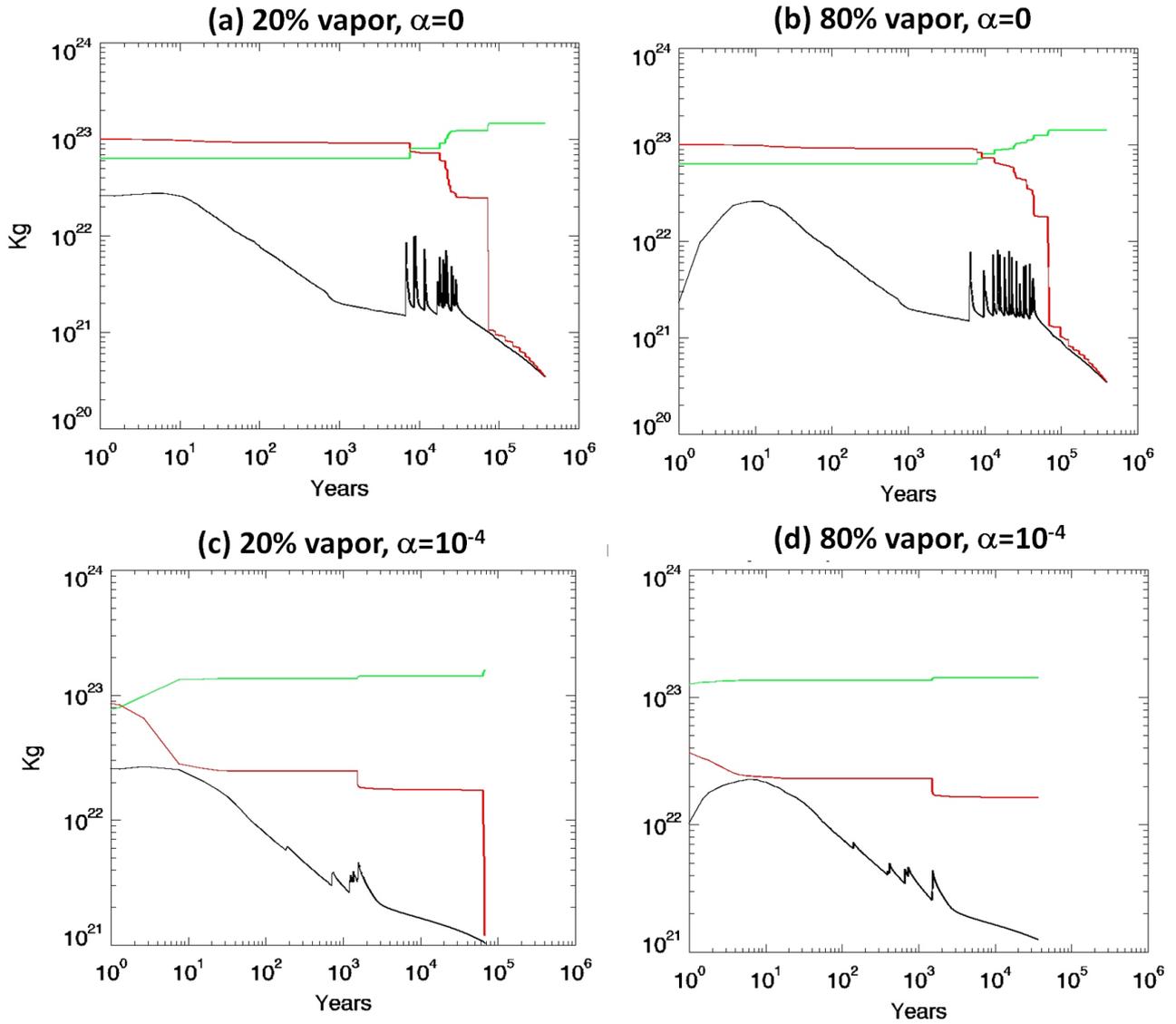

**Figure 8**

Evolution of the protolunar disk's mass starting with initially 20 wt. % or 80 wt. % vapor fraction initially and with either a non viscous vapor disk ($\alpha=0$) or with a viscous vapor disk ($\alpha=10^{-4}$). Red: total disk mass, green : mass fallen on Earth , black: total mass in the gravitationally unstable phase of the disk (that can ultimately be assembled into the Moon). See main text for details.

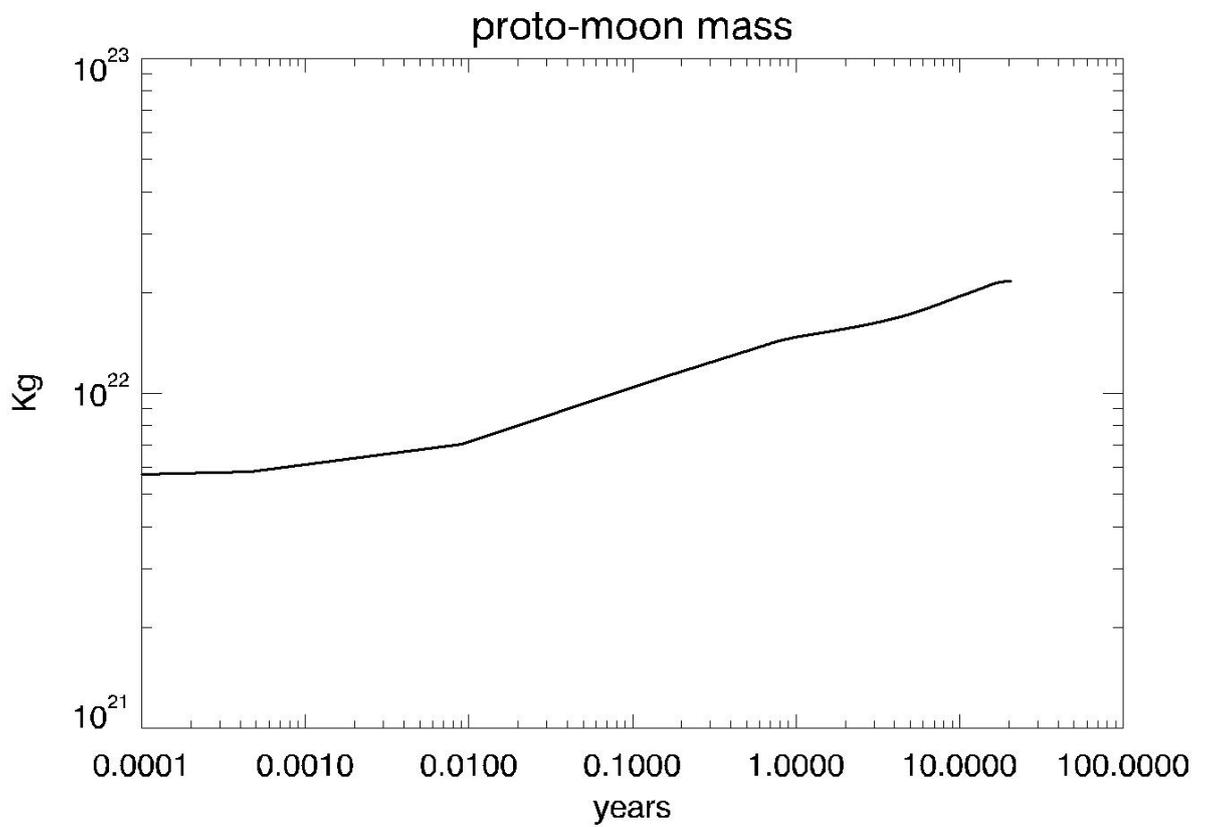

**Figure 9**: Mass of a growing proto-moon assuming that all material spreading outside the Roche Limit is instantaneously incorporated in a proto-moon. All the incorporated material is in liquid form.

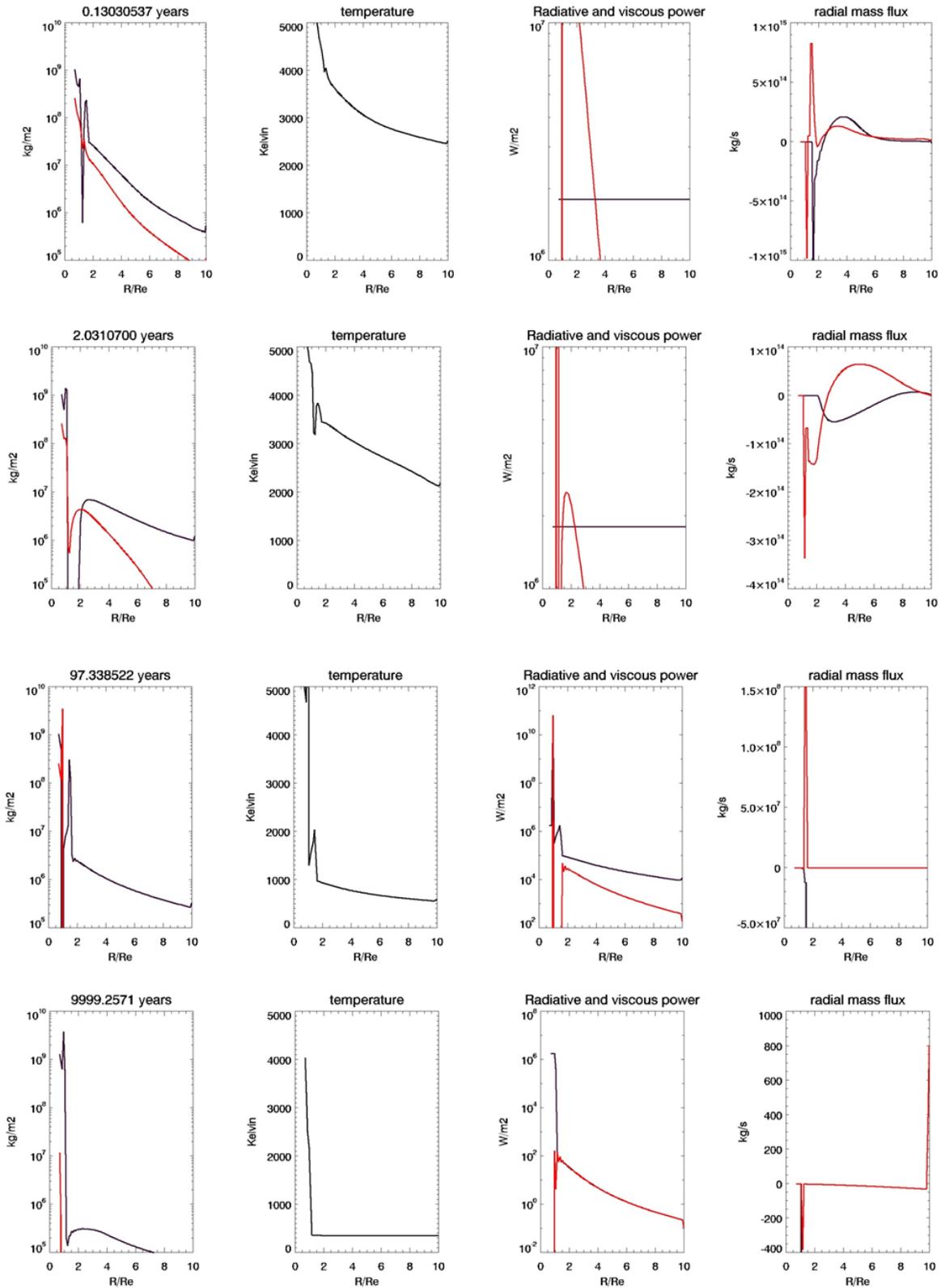

**Figure 10**

Evolution of a disk initially with 20 wt. % vapor initially and for a turbulent gas with $\alpha=0.0001$. Same legend as Figure 3.

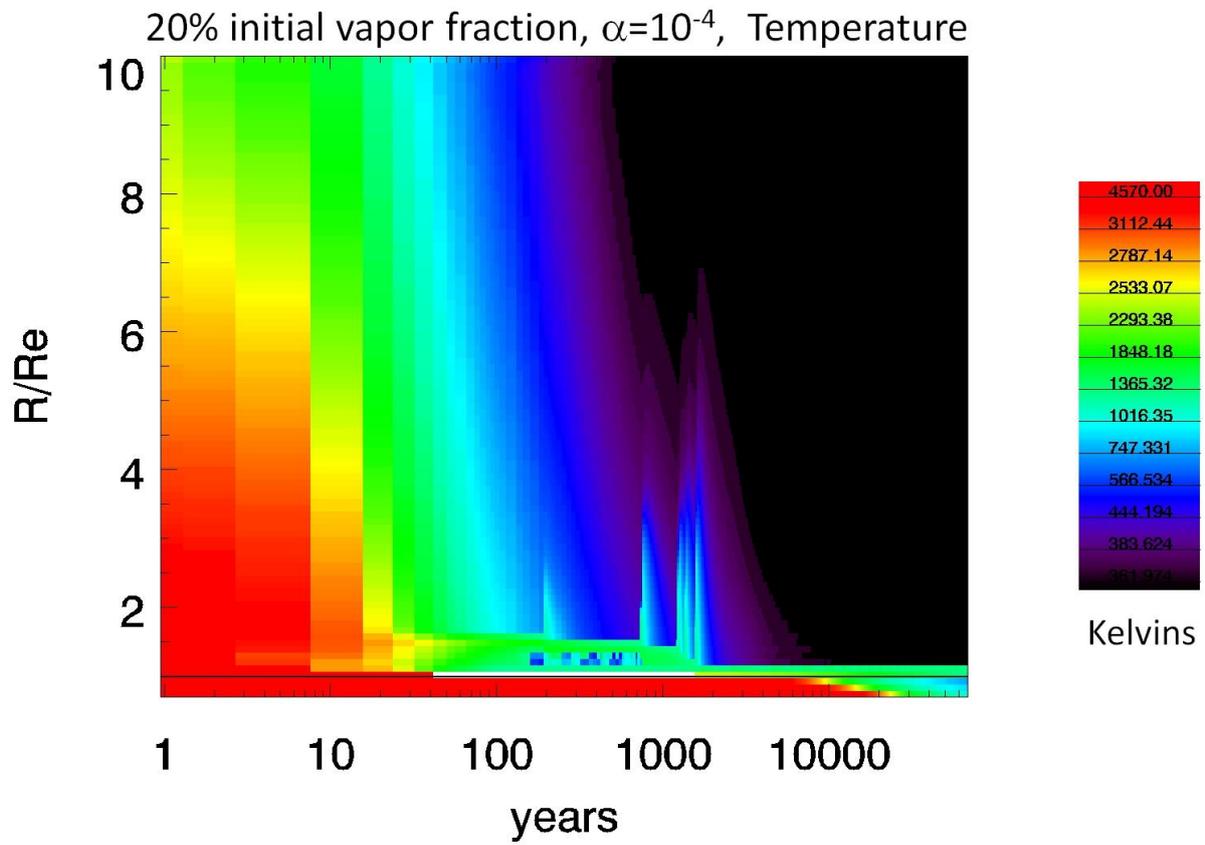

**Figure 11**

Temperature evolution of the disk with 20 wt. % vapor initially, with a turbulent coefficient $\alpha=0.0001$ for the gas phase.

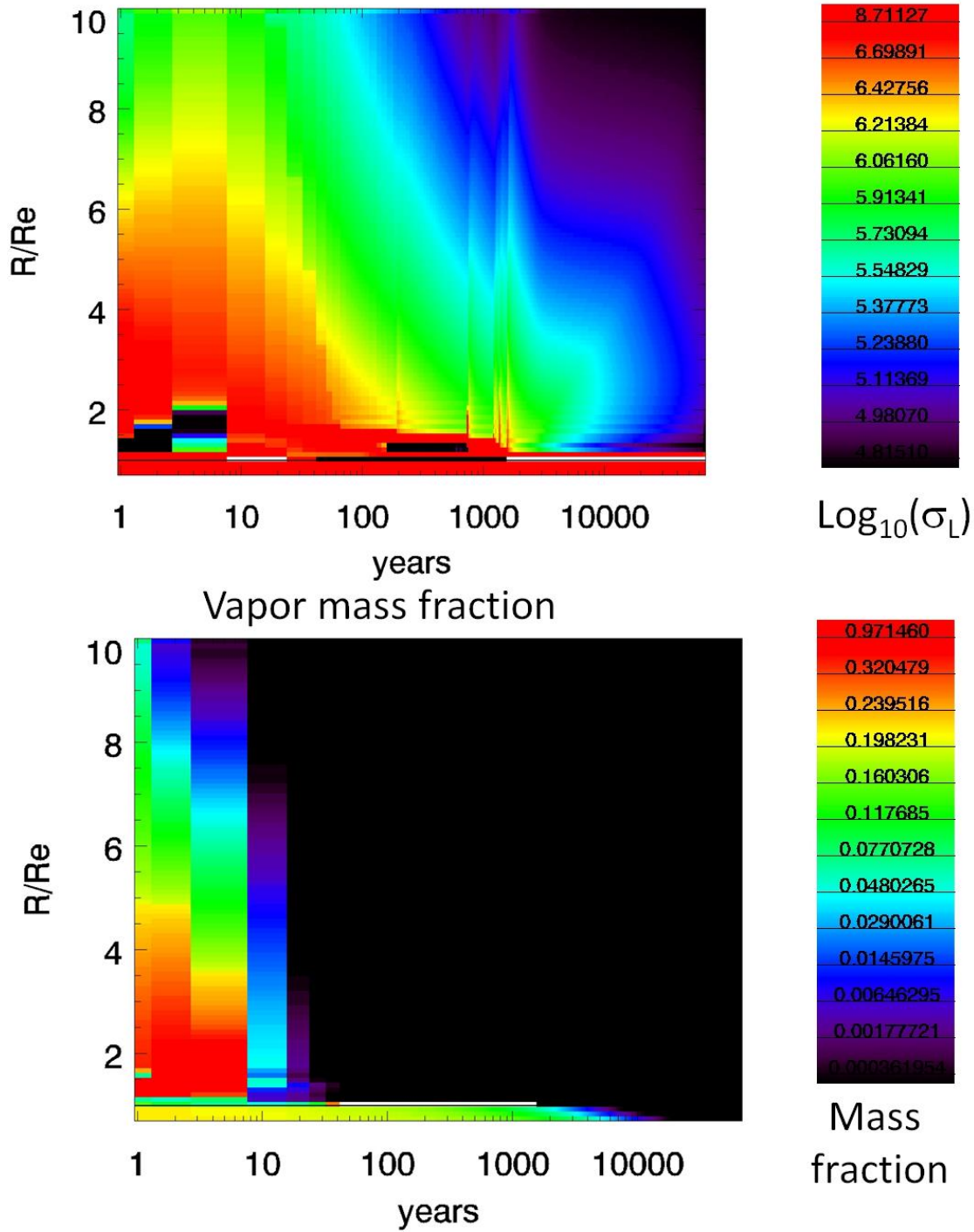

**Figure 12**

Time evolution of a disk with initially 20 wt. % vapor mass fraction and a viscous vapor with a coefficient $\alpha=0.0001$. Top: surface density of the condensed phase, bottom: vapor mass fraction.

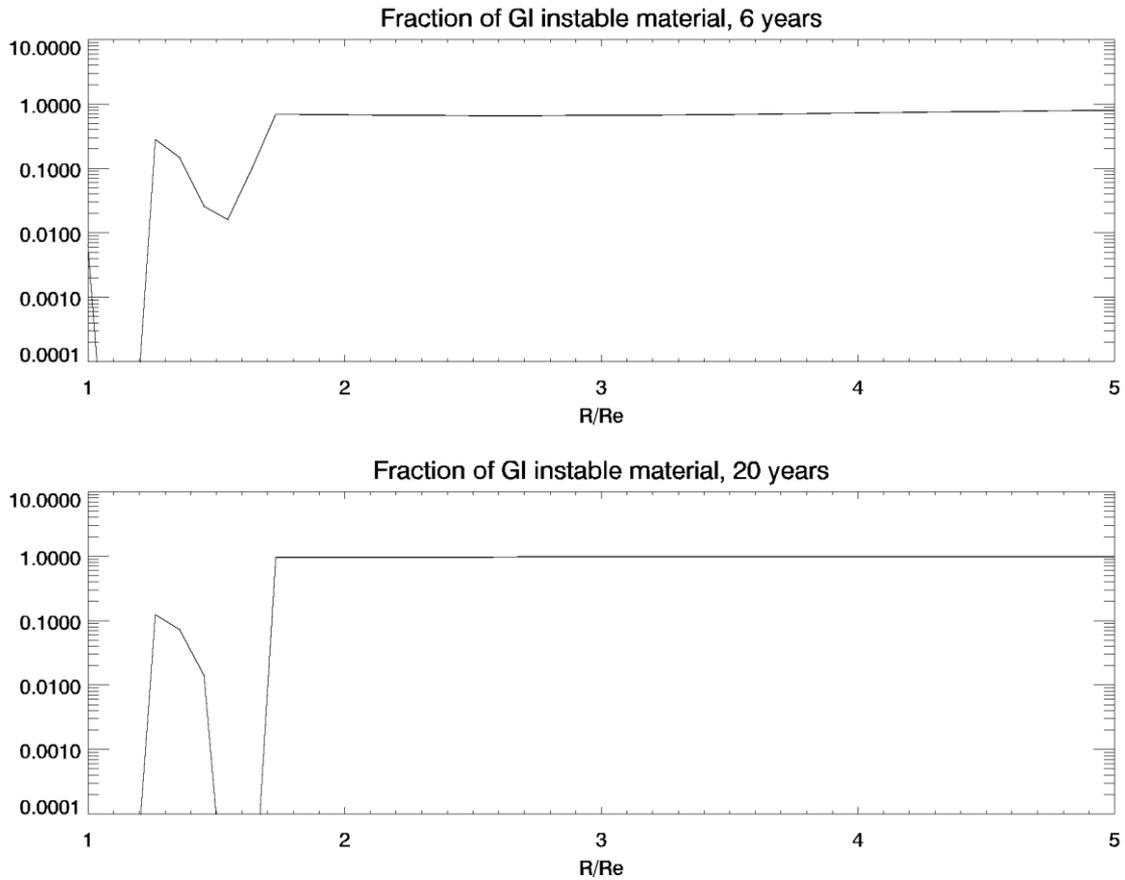

**Figure 13**

Local mass fraction of gravitationally unstable material in a disk subject to Kelvin Helmoltz instability starting initially 20 wt. % vapor, $\alpha=0$ and assuming that Kelvin Helmoltz turbulence mixes vapor and liquid (see section 5.2 for the calculation details). Beyond $R_L$ (1.7$R_e$) the disk is always found to be gravitationally unstable because it is always the case for the liquid phase. Below $R_L$ the KH instability mixes a fraction of the vapor layer with the liquid layer.

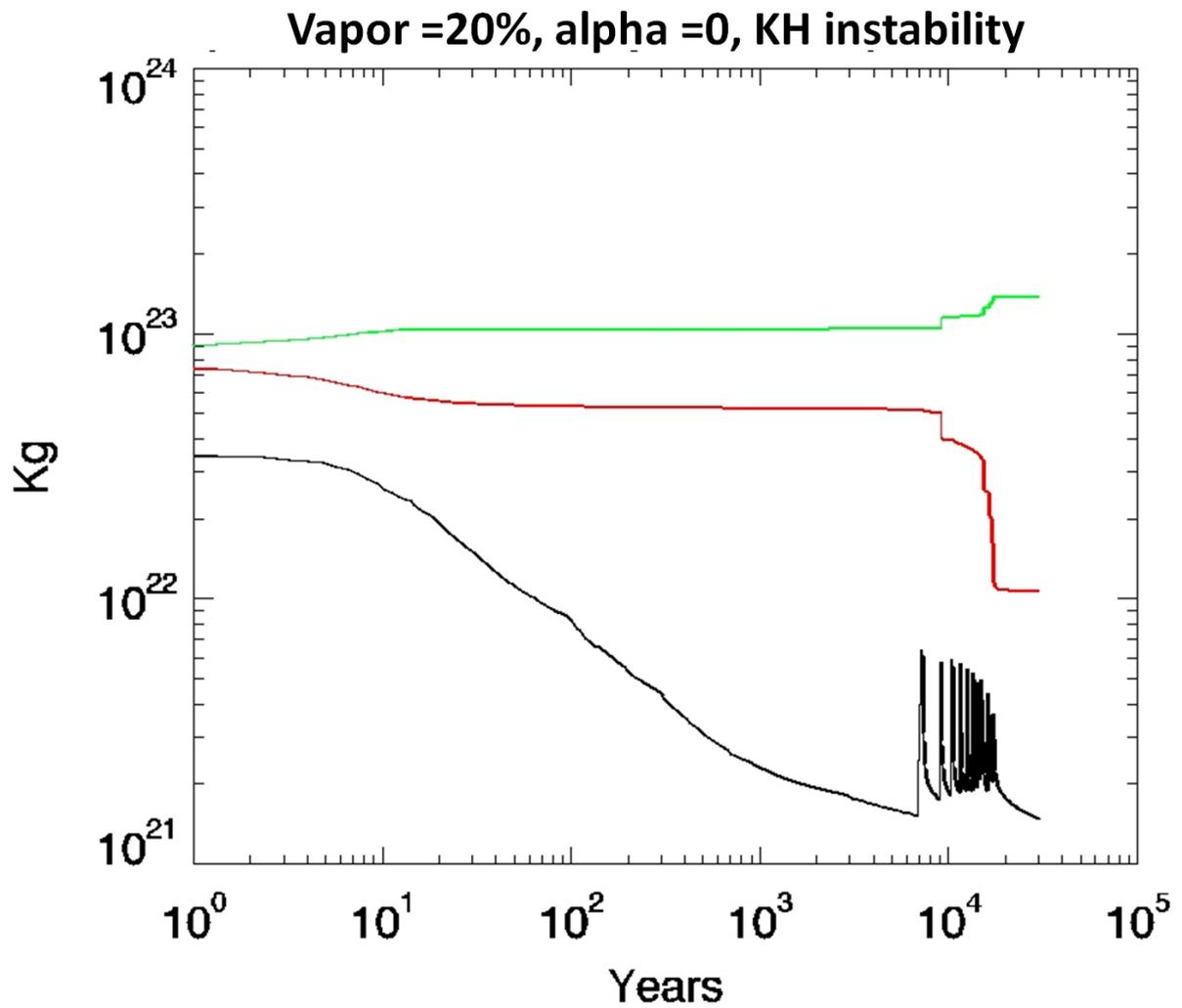

**Figure 14**

Evolution of the protolunar disk mass starting initially 20 wt. % vapor $\alpha$=0 and assuming that Kelvin Helmotz turbulence mixes vapor and liquid (see section 5.2 for the calculation details). Red : total disk mass, green : mass fallen on Earth, black: total mass in the gravitationally unstable phase of the disk that can ultimately be assembled into the Moon.

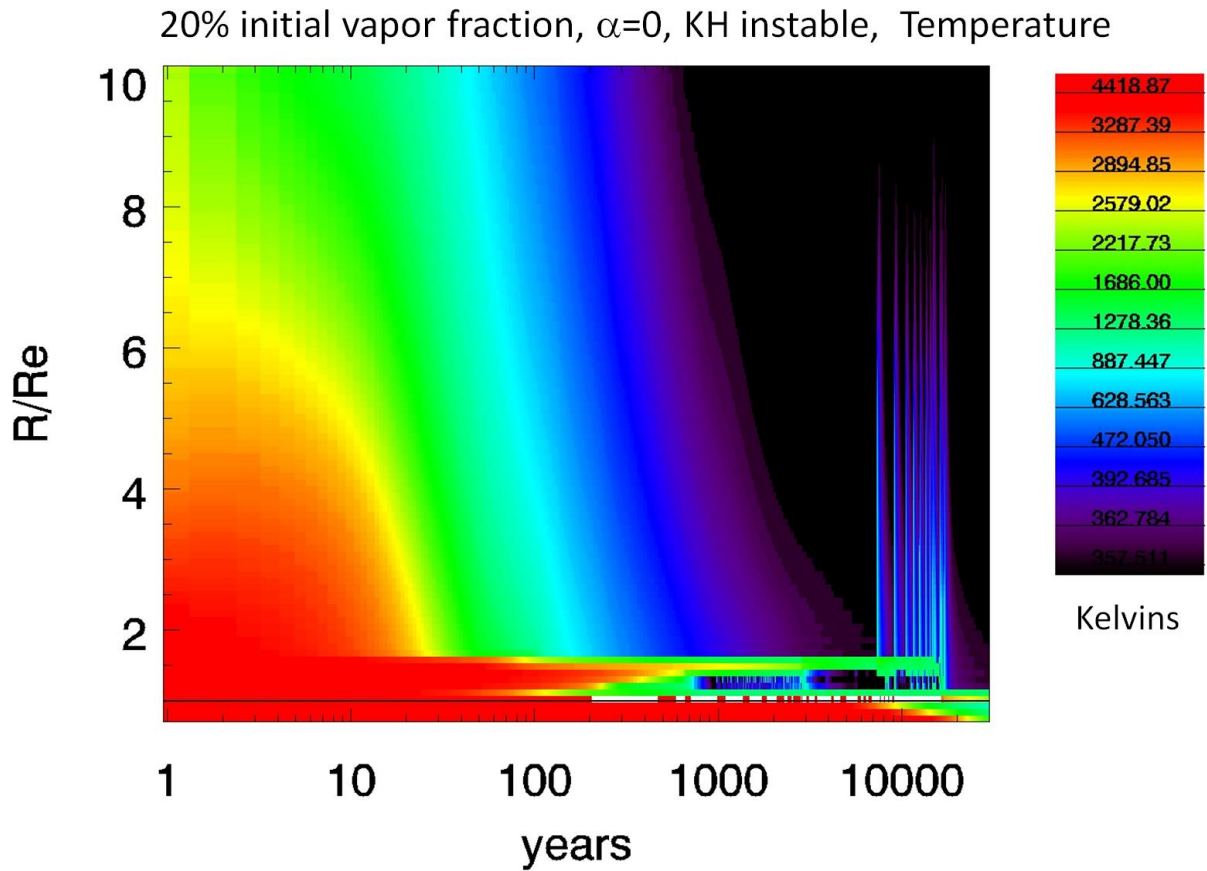

**Figure 15**

Temperature evolution of a disk with initially 20 wt. % vapor $\alpha=0$ and assuming that Kelvin Helmholtz turbulence mixes vapor and liquid (see section 5.2 for the calculation details).

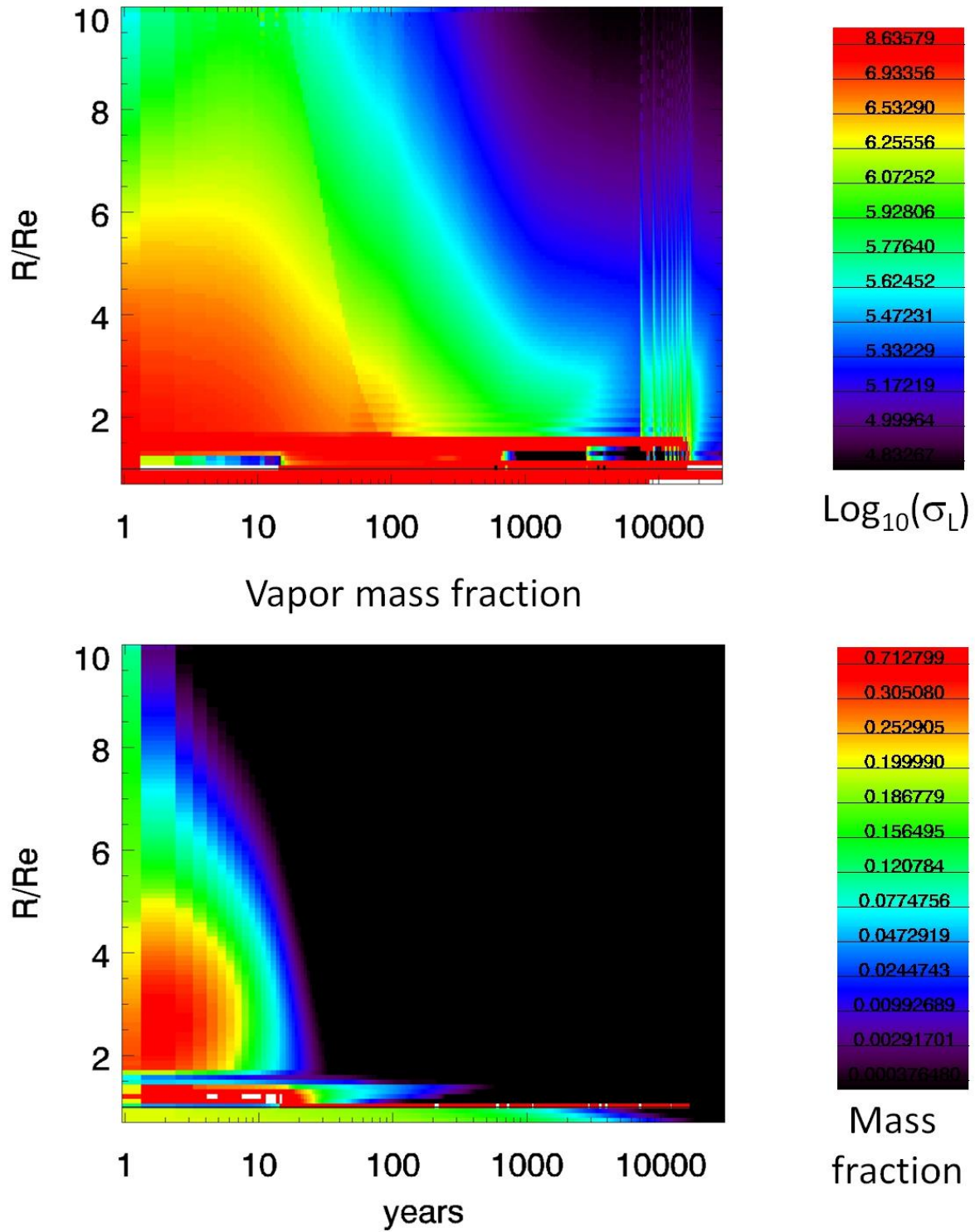

**Figure 16**

Evolution of a disk with initially 20 wt. % vapor mass α=0 and assuming that Kelvin Helmholtz turbulence mixes vapor and liquid (see section 5.2 for the calculation details). Top: surface density of the condensed phase, bottom: vapor mass fraction.

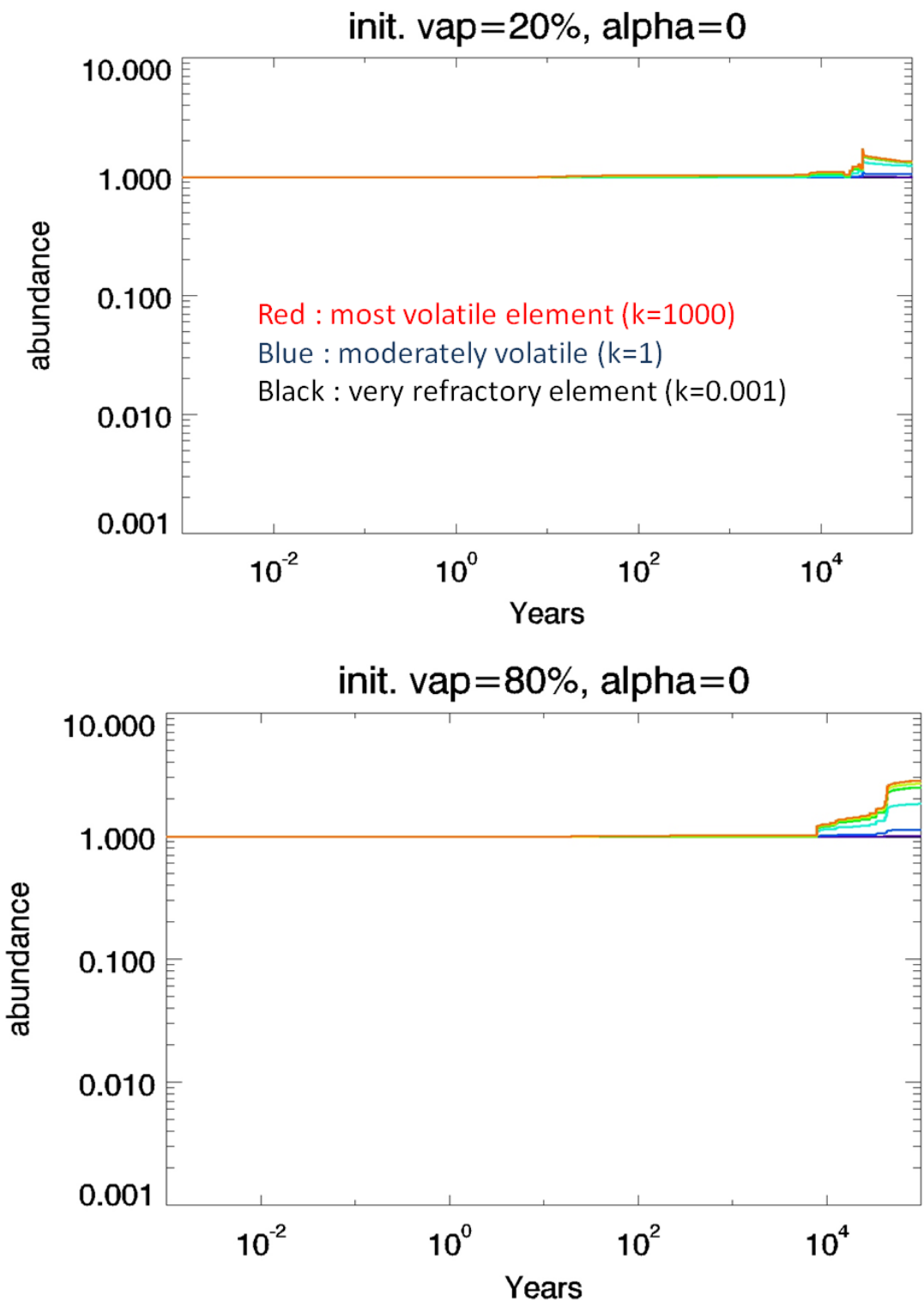

**Figure 17**: Mass fraction of elements with different level of volatility (according to the K parameter, defined in section 4) with respect to the most refractory elements (K=10$^{-3}$). Different colors stand for different levels of volatility : black K=10$^{-3}$, violet : K=10$^{-2}$, dark blue K=10$^{-1}$, light blue K=1, green K=10, yellow: K=10$^2$, red: K=10$^3$. Here the non-viscous vapor disk case is presented for 20% initial vapor fraction (top) and for 80 wt. % initial vapor fraction (bottom).

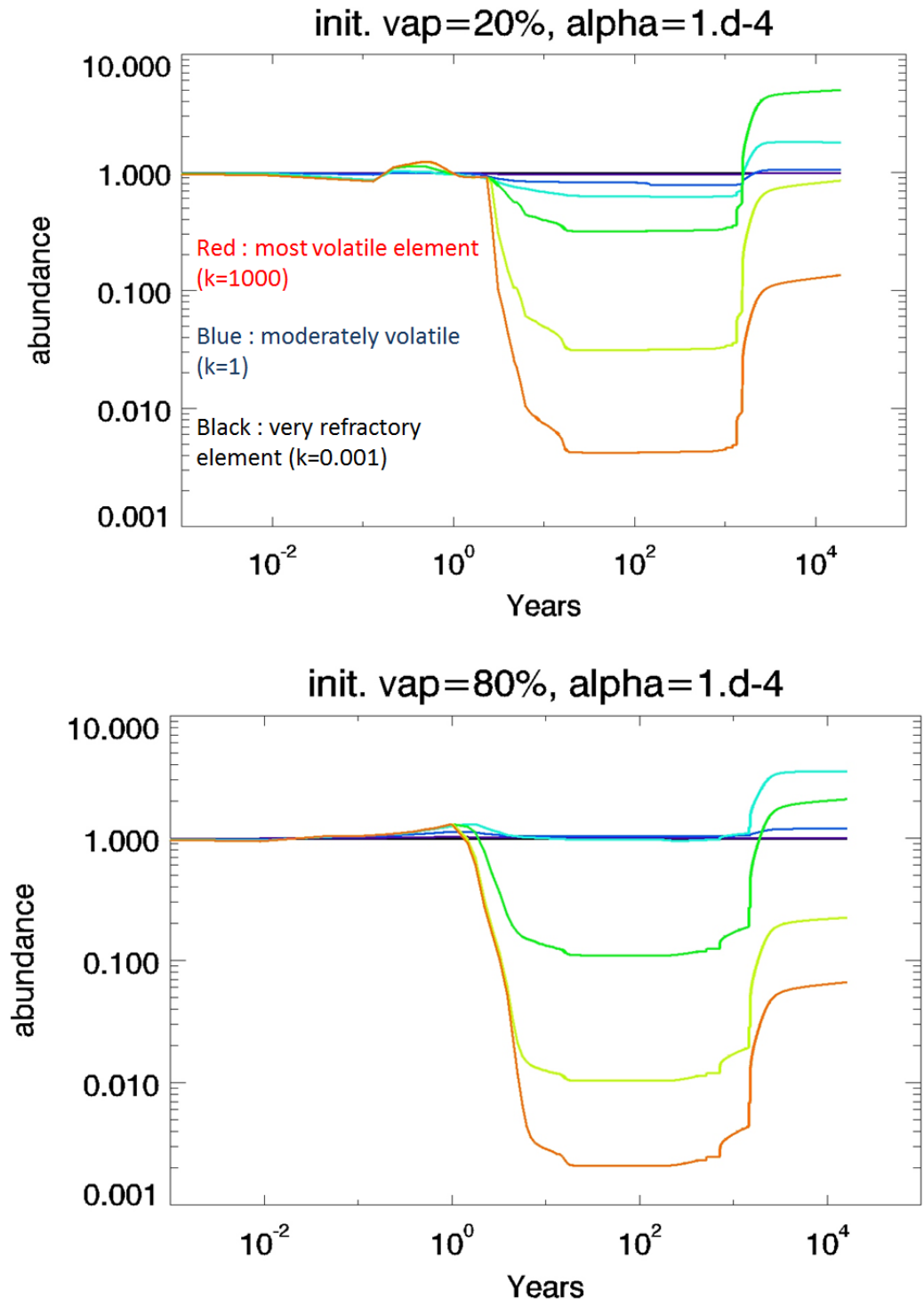

**Figure 18** : Mass fraction of elements with different level of volatility (according to the K parameter, defined in section 4) with respect to the most refractory elements (K=10$^{-3}$). Different colors stand for different levels of volatility : black K=10$^{-3}$, violet : K=10$^{-2}$, dark blue K=10$^{-1}$, light blue K=1, green : K=10, yellow: K=10$^{2}$, red: K=10$^{3}$. Here the viscous vapor disk case is presented with $\alpha$=10$^{-4}$ with (top) 20 wt. % initial vapor fraction and (bottom) 80 wt. % initial vapor fraction

# Appendix A : Comparing isothermal and Clapeyron structures

Here we compare the structure and energy of a three-layer disk, composed of a condensed central disk overlaid by two layers of a gaseous atmosphere, in the case where the atmosphere is isothermal or is on the Clapeyron curve.

In the case where the atmosphere is on the Clapeyron curve, the three-layer structure might be conserved if the condensed droplets rapidly rains out from the atmosphere on to the central condensed disk by gravity.

In the case where the atmosphere is isothermal at temperature $T_s$, the surface density of the vapor atmosphere $\sigma_v$ is calculated by two successive integration of the hydrostatic equation from z=d, the height of the midplane layer to infinity:

$$\frac{dP}{P} = -\frac{\mu}{RT_s}\omega_k^2 z dz$$

where we assume a perfect gas, i.e. $\rho_v = \frac{P\mu}{RT}$, hence $\sigma_v$ is given by:

$$\sigma_v = 2\rho_{vs}\exp\left(\frac{d^2}{2H_{Ts}^2}\right)\int_d^\infty \exp\left(-\frac{z^2}{2H_{Ts}^2}\right)dz \qquad \text{(Equation A1)}$$

where $\rho_{vs}$ is the density of the vapor at temperature $T_s$ and pressure $P_s$ at the phase transition, d is half the thickness of the midplane condensed layer and $H_{Ts}$ is the local pressure scale height at temperature $T_s$:

$$H_{Ts}^2 = \frac{RT_s}{\mu\omega_k^2} \qquad \text{(Equation A2)}$$

Hence the surface density of the vapor layer is is

$$\sigma_v = \rho_{gs}\sqrt{2\pi}H_{Ts}\exp\left(\frac{d^2}{2H_{Ts}^2}\right)erfc\left(\frac{d}{\sqrt{2}H_{Ts}}\right) \qquad \text{(Equation A3)}$$

which, for $d \ll H_{Ts}$, resumes to

$$\sigma_v \approx \rho_{vs}\sqrt{2\pi}H_{Ts} - 2\rho_{vs}d \approx \rho_{vs}\sqrt{2\pi}H_{Ts} \qquad \text{(Equation A4)}$$

In that case the internal energy per unit surface of the atmosphere is simply

$$e_v = C_{vg}T_s\sigma_v \approx \rho_{vs}C_{vg}T_s\sqrt{2\pi}H_{Ts} \qquad \text{(Equation A5)}$$

In the case where the atmosphere is on the Clapeyron curve, the surface density of the vapor atmosphere $\sigma_v$ is still calculated by integration of the hydrostatic equation between the height of the midplane layer d, where $P=P_s$ and $T=T_s$, the pressure and temperature of phase transition, and infinity, but using the Clapeyron equation for the temperature:

$$\frac{dP}{P\ln(P/P_0)} = -\frac{\mu}{RT_0}\omega_k^2 z\,dz$$

$$\Rightarrow \ln\left(\frac{P(z)}{P_0}\right) = -\frac{T_0}{T_s}\exp\left(\frac{z^2-d^2}{H_{T_0}^2}\right)$$

where

$$H_{T_0}^2 = \frac{RT_0}{\mu\omega_k^2}$$

is large since $T_0$ is large. For $z \ll H_{T_0}$, where most of the vapor mass is located, we have, by developing $\exp\left(\frac{z^2-d^2}{H_{T_0}^2}\right)$ around 0:

$$\ln\left(\frac{P(z)}{P_0}\right) = -\frac{T_0}{T_S}\left(1 + \frac{z^2-d^2}{H_{T_0}^2}\right)$$

$$\Rightarrow \frac{P(z)}{P_0} \approx \exp\left(-\frac{T_0}{T_s}\right)\exp\left(-\frac{z^2-d^2}{H_{Ts}^2}\right)$$

$$\Rightarrow P \approx P_s \exp\left(-\frac{z^2-d^2}{H_{Ts}^2}\right)$$

$$\Rightarrow \rho_v \approx P_s \frac{\mu}{RT}\exp\left(-\frac{z^2-d^2}{H_{Ts}^2}\right)$$

$$\Rightarrow \rho_v \approx P_s \frac{\mu}{RT_0}\ln\left(\frac{P_0}{P}\right)\exp\left(-\frac{z^2-d^2}{H_{Ts}^2}\right)$$

$$\Rightarrow \rho_v \approx \rho_s \frac{T_s}{T_0}\left(\frac{T_0}{T_s} + \frac{z^2-d^2}{H_{Ts}^2}\right)\exp\left(-\frac{z^2-d^2}{H_{Ts}^2}\right)$$

$$\Rightarrow \rho_v \approx \rho_s\left(1 + \frac{z^2-d^2}{H_{T_0}^2}\right)\exp\left(-\frac{z^2-d^2}{H_{Ts}^2}\right) \approx \rho_s \exp\left(-\frac{z^2-d^2}{H_{Ts}^2}\right)$$

which gives $\sigma_v$ by integration the surface density of the atmosphere

$$\sigma_v = 2\rho_s \int_d^\infty \exp\left(-\frac{z^2-d^2}{H_{Ts}^2}\right)dz = H_{Ts}\rho_s\sqrt{2\pi} - 2\rho_s d \approx H_{Ts}\rho_s\sqrt{2\pi}$$

which is equivalent to the one for an isothermal atmosphere at temperature $T_s$.

Now the internal energy $e_v$ of the vapor layer is given by

$$e_v = 2\int_d^\infty C_{vg} T dz$$

$$\Rightarrow e_v = 2C_{vg} \int_d^\infty \frac{T_0}{\ln(P_0/P)} dz$$

$$\Rightarrow e_v = 2C_{vg} T_0 \int_d^\infty \frac{T_s}{T_0}\left(1 + \frac{z^2 - d^2}{H_{T_0}^2}\right)^{-1} dz$$

$$\Rightarrow e_v = 2C_{vg} T_s \left[\frac{H_{T_0}^2}{\sqrt{H_{T_0}^2 - d^2}} \arctan\left(\frac{z}{\sqrt{H_{T_0}^2 - d^2}}\right)\right]_d^\infty$$

$$\Rightarrow e_v = 2C_{vg} T_s \frac{H_{T_0}^2}{\sqrt{H_{T_0}^2 - d^2}} \left[\frac{\pi}{2} - \arctan\left(\frac{d}{\sqrt{H_{T_0}^2 - d^2}}\right)\right]$$

$$\Rightarrow e_v \approx C_{vg} T_s \pi H_{T_0}$$

if d<<$H_{T0}$.

Hence, if the gas atmosphere lies on the Clapeyron curve, the energy stored in the gas is larger by a factor $\sqrt{\frac{\pi T_0}{2T_s}}$ than if the gas is on an isotherm at the temperature $T_s$.

# SUPPLEMENTARY ONLINE MATERIAL

## 1. Detailed evolution of a disk with starting with 80 wt. % vapor

**1.1 Non viscous vapor disk**

We consider the case of a disk with 80 wt. % vapor initially, and with no turbulent viscosity ($\alpha$=0). Such a high vapor fraction is encountered in disks formed after a high velocity impact like in the model of Cuk and Stewart (2012) or after collisions of two sub-Earths (Canup 2012) or in grazing impacts (Reufer et al. 2012). The initial disk evolution is presented in Figure 1 of the supplementary online material. In the first 10 years the disk cools down and becomes mostly liquid (Figure 1.a, 1.b, 1.c of the supplementary online material), and the material rapidly accumulates below $R_L$ like in the 20 wt. % vapor case. On a long term evolution the disk cools down exterior to $R_L$ (that contains less than 1% of the disk mass) in about 100 years while the hot and compact liquid disk interior to $R_L$ remains hot for $10^4$ years. Interestingly the hot and compact liquid disk between $R_\oplus$ and $R_L$ contains in average more vapor than in the 20% vapor case: the vapor mass fraction is >40% for the first 500 years (Figure 2 of the supplementary online material bottom panel). Pahlevan and Stevenson (2007) suggested that a vapor rich disk lasting for at least 1000 years may allow a compositional equilibration of the disk with the Earth's. It seems that a disk starting with a high vapor fraction may offer such an adequate context. Indeed, following the giant impact, a significant fraction of the Earth's mantle is likely to be in a magma ocean state which would help equilibration with the disk. In particular, recent experimental studies at high pressure and temperature emphasize the need of a deep initial magma ocean to explain metal equilibration during core formation in the Earth (Siebert et al, 2012). Cooling of the magma ocean would also occur on a longer timescale than the disk lifetime (Abe, 1997; Elkins-Tanton, 2012; Lebrun et al, 2013) and thus does not limit equilibration of species. After $10^4$ years evolution, heat bursts are also visible and help maintaining the disk hot up to about $10^{5'}$ years. After $10^5$ years the disk is entirely in solid form (Figure 1 of the supplementary online material, row 4) and most of its mass has fallen onto the Earth (Figure 3 top of the supplementary online material).

**1.2 Viscous vapor disk**

The evolution of a disk starting with 80 wt.% vapor in mass, and with a viscous vapor assuming $\alpha=10^{-4}$ is presented in Figure 4 of the supplementary online material. The evolution is very similar to the 80 wt. % vapor case discussed in section 3.4 of the main text.

## vapor f.=80%, non-turbulent vapor

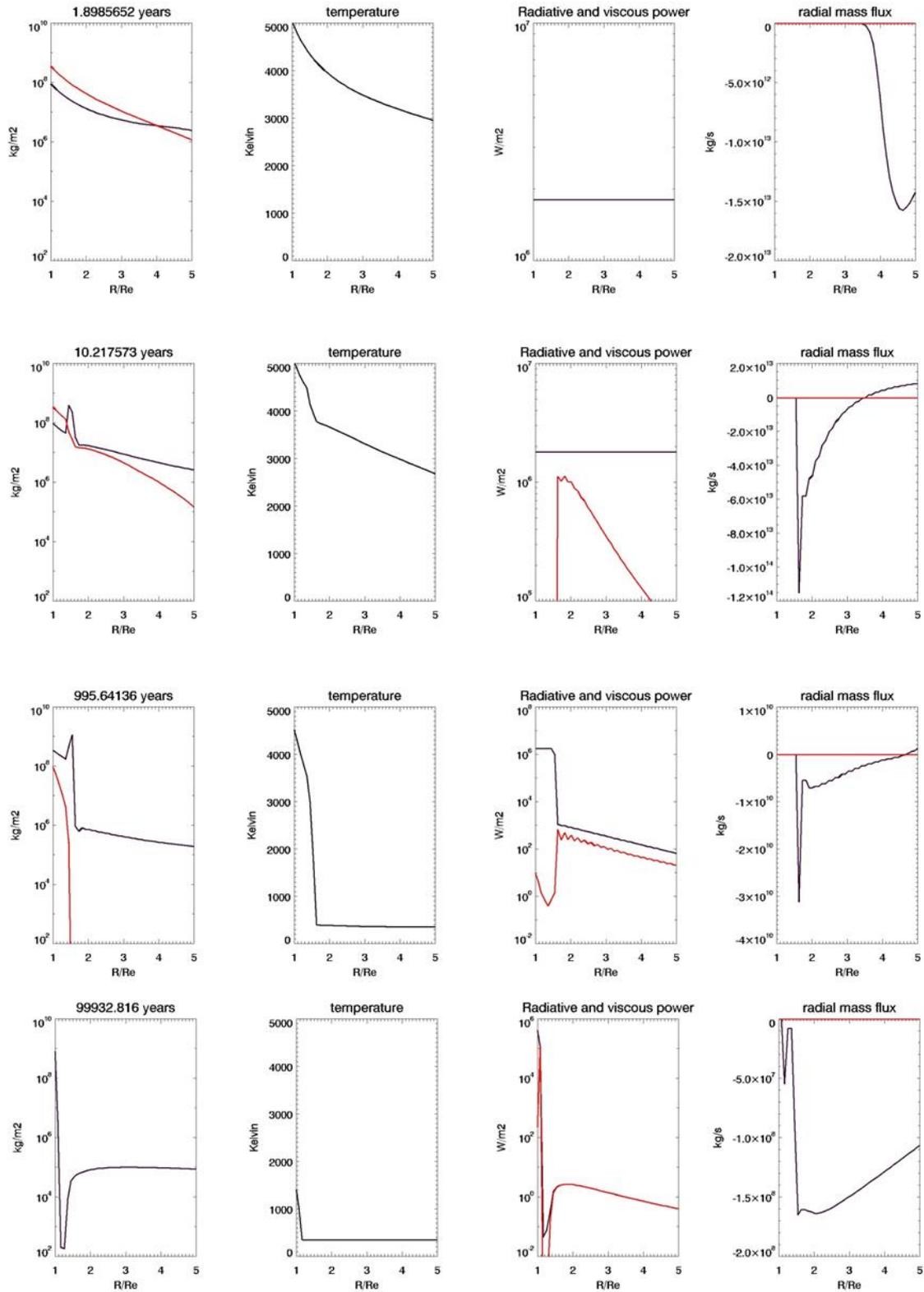

**Supplementary Figure 1 :** Time evolution of a protolunar disk starting with 80 wt. % initial vapor and in the case of a non turbulent vapor.

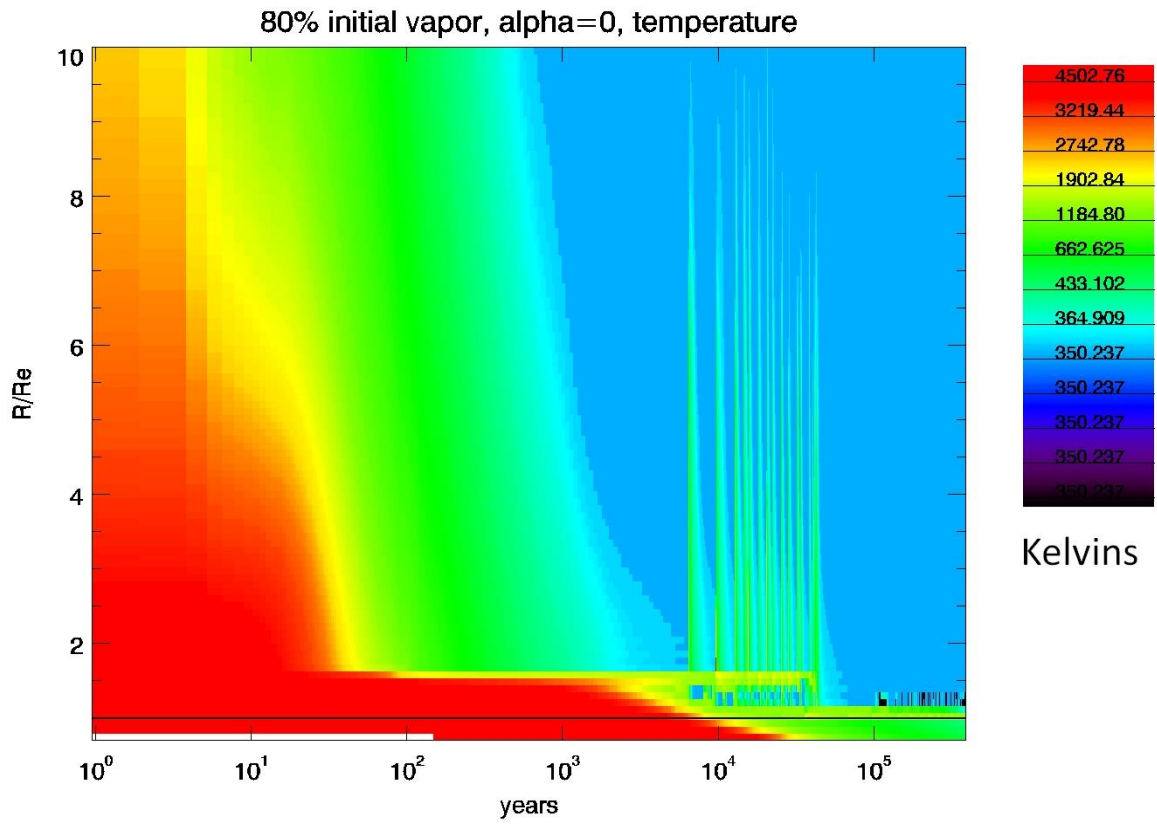

**Supplementary Figure 2.**

Temporal and spatial evolution of the temperature for a disk starting with 80 wt. % vapor mass fraction.

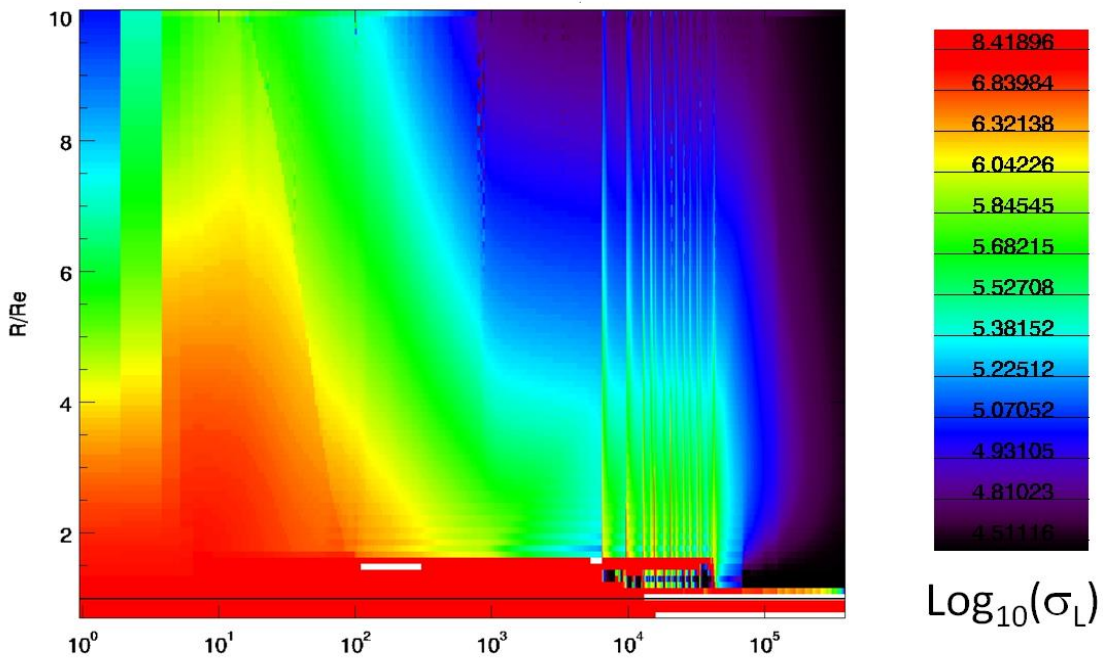

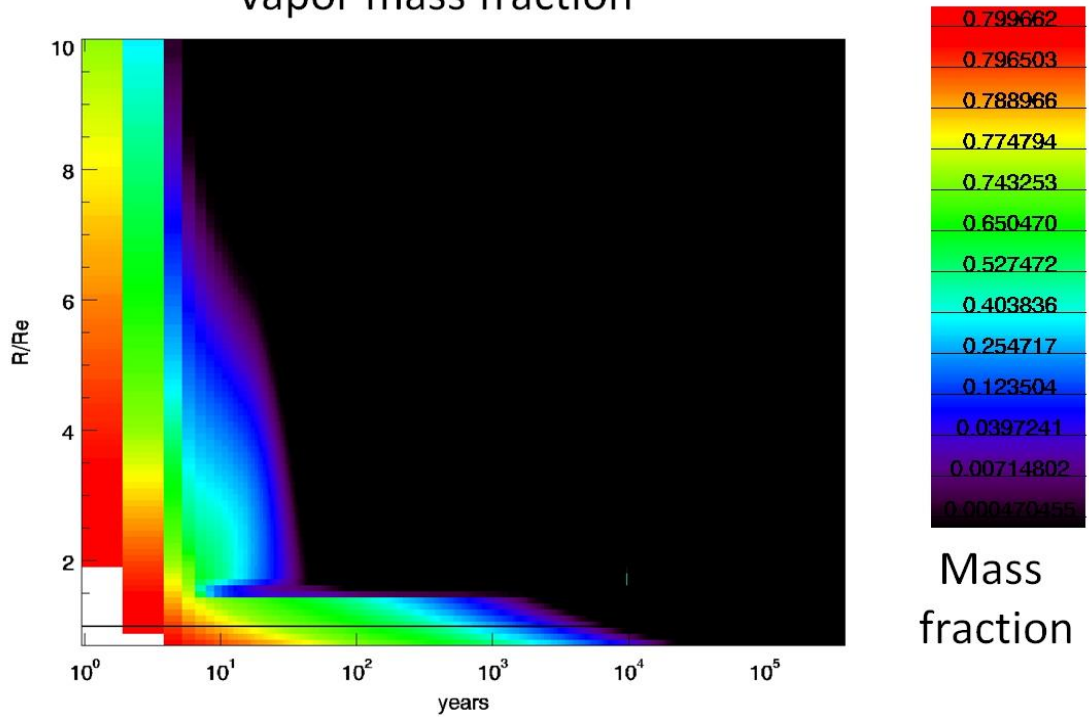

**Supplementary Figure 3.**

Time evolution of a disk starting with 80 wt. % vapor mass fraction and a non-viscous vapor. Top: surface density of the condensed phase, bottom: vapor mass fraction.

## vapor fraction=80%, $\alpha=10^{-4}$

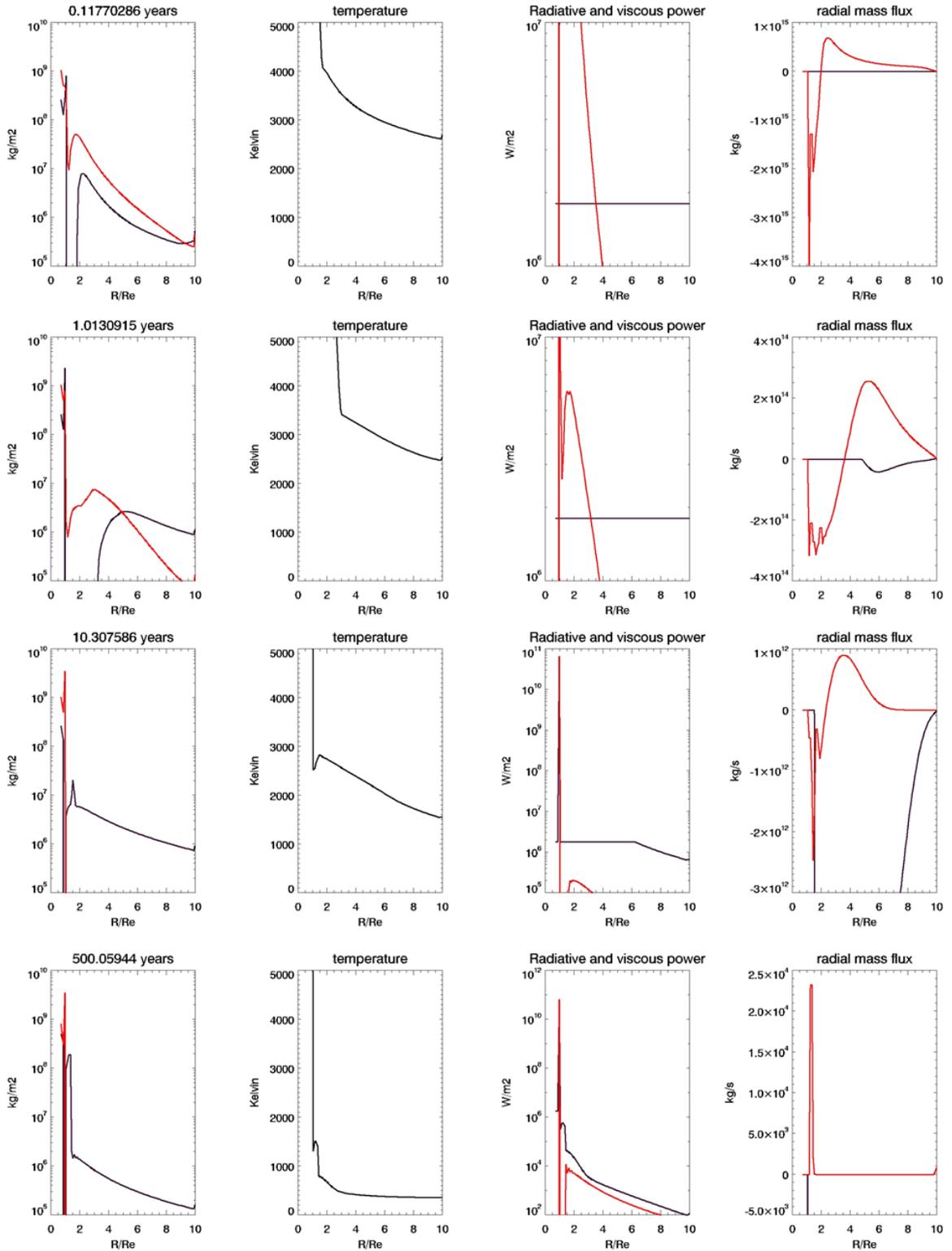

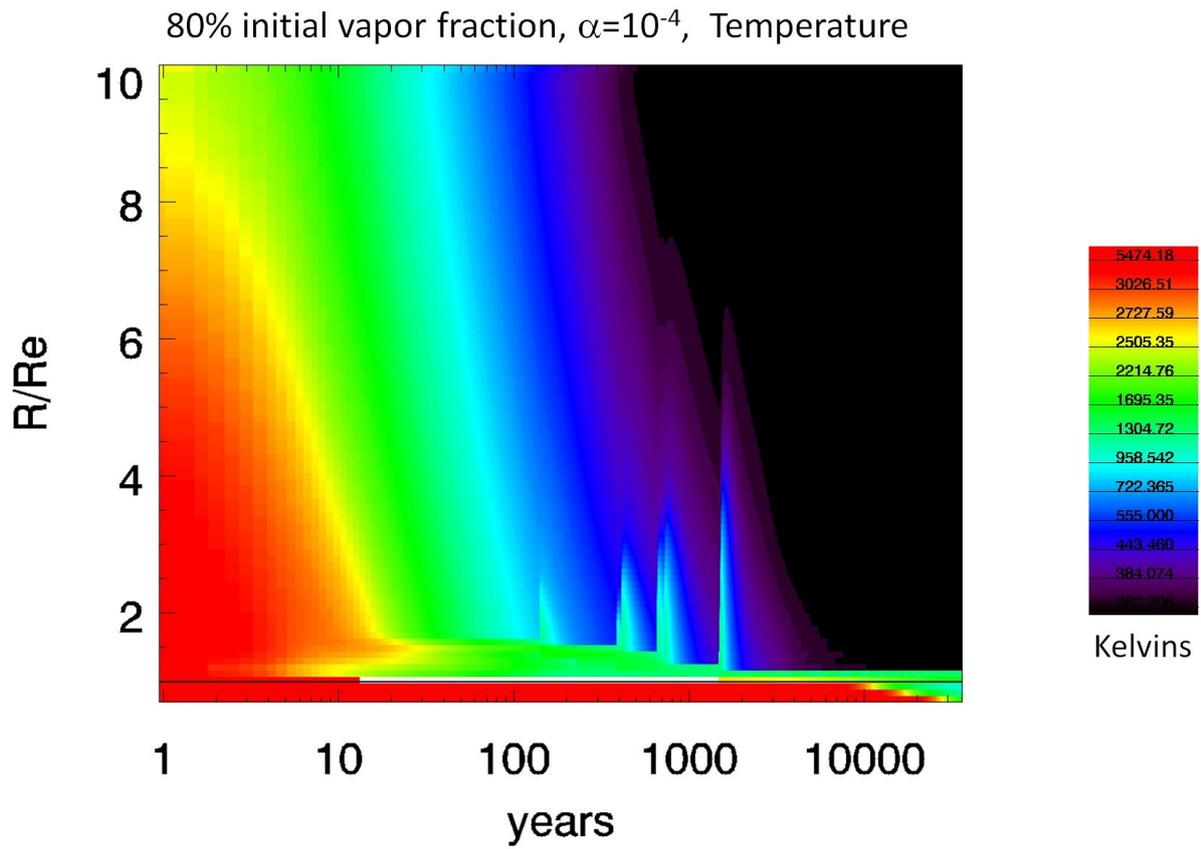

**Supplementary Figure 5.**

Temperature evolution of the disk starting with 80 wt. % vapor, with a turbulent coefficient $\alpha=0.0001$ for the gas phase.

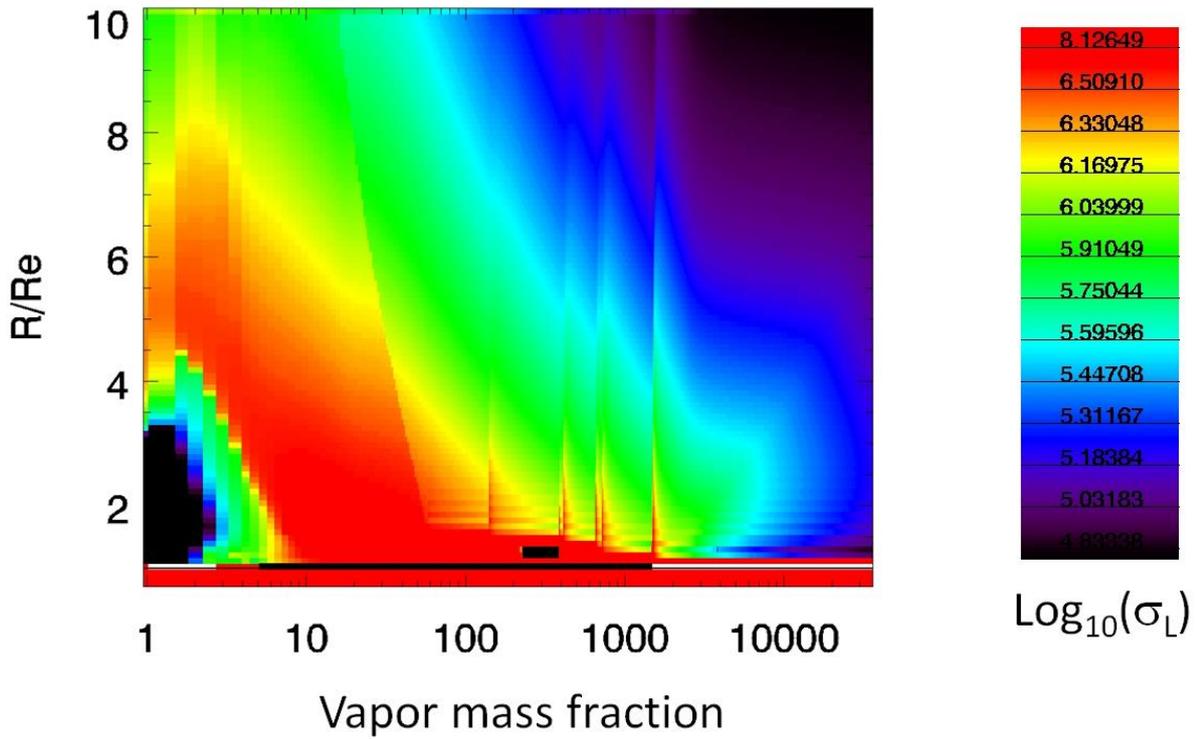

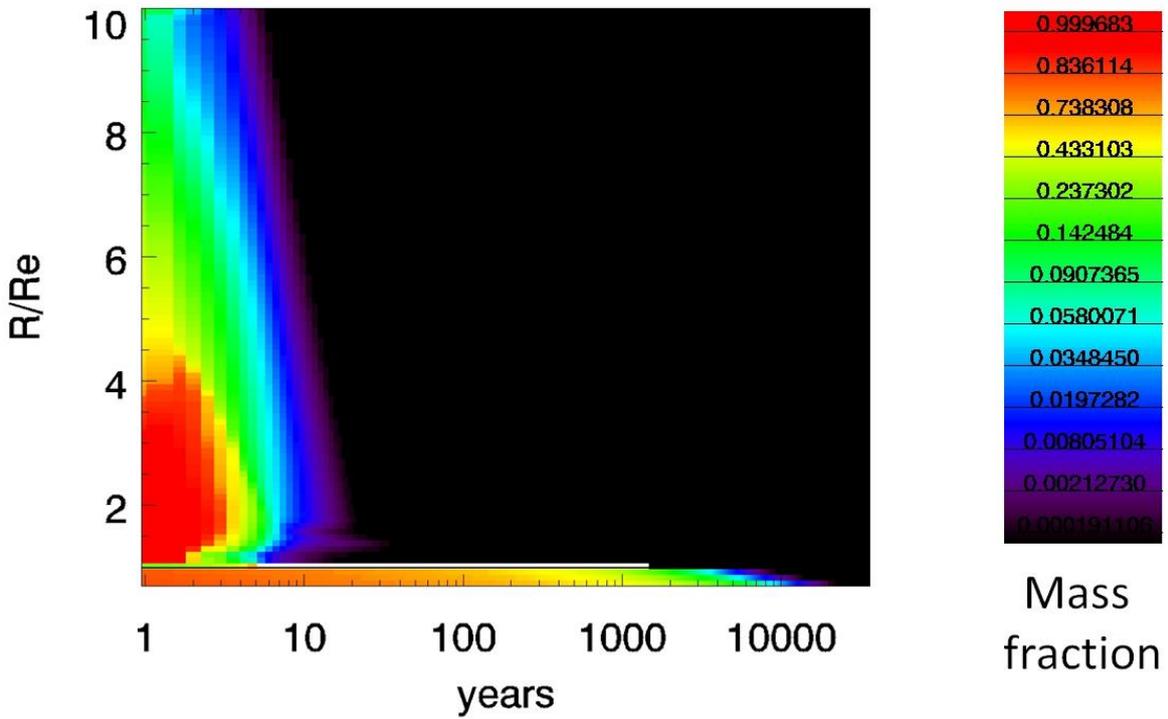

**Supplementary Figure 6.**

Time evolution of a disk with initially 80 wt. % vapor mass fraction and a viscous vapor with coefficient $\alpha$=0.0001. Top: surface density of the condensed phase, bottom: vapor mass fraction.

## 2. Exploring Different disk masses

We performed an additional suite of simulations with more massive initial disks, starting with about 4 lunar masses, to explore the effect of disk mass on the disk evolution. This is about the upper value of disk masses reported in the literature (see e.g. Canup 2004, 2012 or Cuk & Stewart 2012, Reufer et al., 2012). The disk mass evolution is displayed in Figure 7 of the supplementary online material. The evolution is qualitatively the same as in the 1.5 lunar mass case (sections 3.3 and 3.4 of the main text): the mass in the unstable phase decreases with time, to the point where burst are triggered when the disk interior to $R_L$ cools down below the solidification temperature. However, the mass stored in the gravitationally unstable disk in the 4 lunar masses case is about the same as in the 1.5 lunar mass case after 100 years only. As a result the disk mass stored in the stable disk below $R_L$ is larger in the 4 lunar masses case. So it seems that even increasing the initial disk mass does not really help to increase the mass stored in the gravitationally unstable disk at the end of the disk evolution. The additional mass simply goes into the hot liquid compact disk and flows down onto the Earth's surface. A somewhat similar result has been found in the study of viscous evolution of Saturn's rings (Salmon et al., 2010) where the feedback between the viscosity and the disk surface density controls the total disk mass over time and results in a gravitationally unstable disk with always the same final mass, such that Q~2 everywhere . Indeed if the disk is very massive, with Q<<1, the viscosity is high and the material flows efficiently inward and piles up inside the stable region below $R_L$. The lower the Q, the higher the mass flux (see section 2). Conversely when the disk surface density is low enough and $Q \geq 2$ the mass flow is reduced (see e.g. Salmon et al., 2010). So there is a feedback between the disk mass and the mass flow forcing the unstable disk to flow inward until it settles in a state with a surface density low enough such that Q~2 everywhere. As the liquid disk is stable below $R_L$ the more massive it is initially, the longer it takes to cool down. This explains the delay (by a factor of 2 to 4) in the onset final solidification compared to the 1.5 lunar mass case (compare Figure 5 of the supplementary online material with Figure 6 of the main paper). The same discussion applies to the case of a turbulent disk and abnormally viscous disk. In conclusion: despite an increase in the disk initial mass, the total mass of the gravitationally unstable disk, that can contribute to the Moon formation, is close to constant, whereas the mass of the hot compact liquid disk (inside $R_L$) increases. When the hot liquid compact disk solidifies, in all cases studied here, the material flows preferentially inward, onto Earth, rather than outward, disfavoring Moon's formation with its present mass.

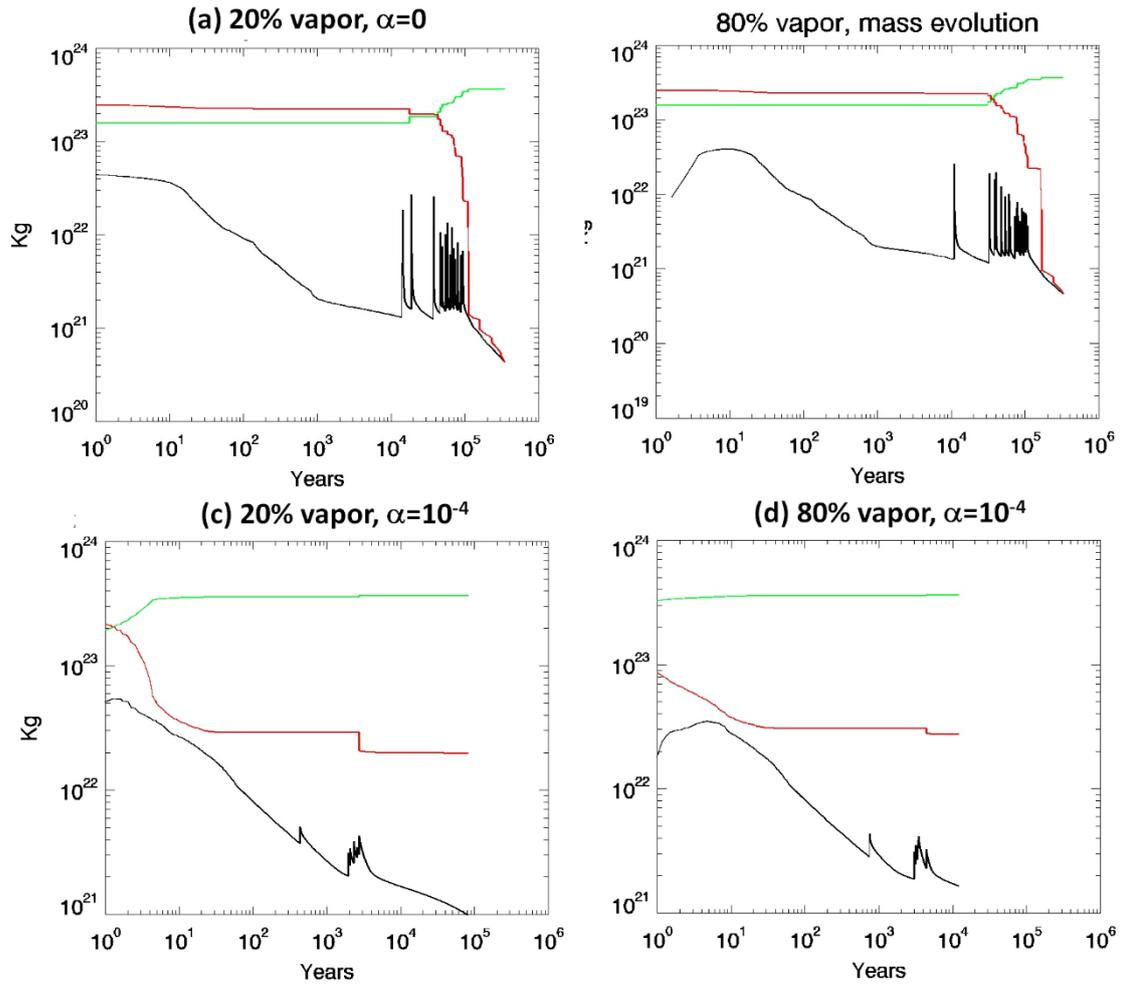

**Supplementary Figure 7**

Time evolution of disks starting with 4 lunar masses of material. Different cases are displayed: disks starting with initially 20% or 80% vapor fraction and with either non viscous vapor layers ($\alpha=0$) or with viscous vapor layers ($\alpha=10^{-4}$). Red line: total disk mass, green line: mass fallen on Earth, black line: total mass in the gravitationally unstable phase of the disk (that can ultimately be assembled into the Moon).

## 3. Disk with a growing proto-moon

We present here the disk' evolution with an ersatz of moon accretion simulated by removing all liquid beyond 3 Earth's Radii (the Roche Limit) and starting with a disk composed of 20 wt. % vapor. Whereas the evolution of the disk's inner regions remains somewhat unaffected (below 2 Earth radii), the disk's outer region (just below the Roche Limit) is significantly affected by the strong outflow induced by the presence of the sharp edge. Note that the disk evolution could be computed only during about 10 years due to very small time-steps imposed by the sharp edge at the Roche Limit. The mass of the proto-moon versus time is displayed in Figure 9 of the main paper while the temperature, vapor fraction and liquid surface density evolutions are displayed in Figures 8 to 10 of the supplementary online material. See also the discussion in section 3.3.3 of the main paper.

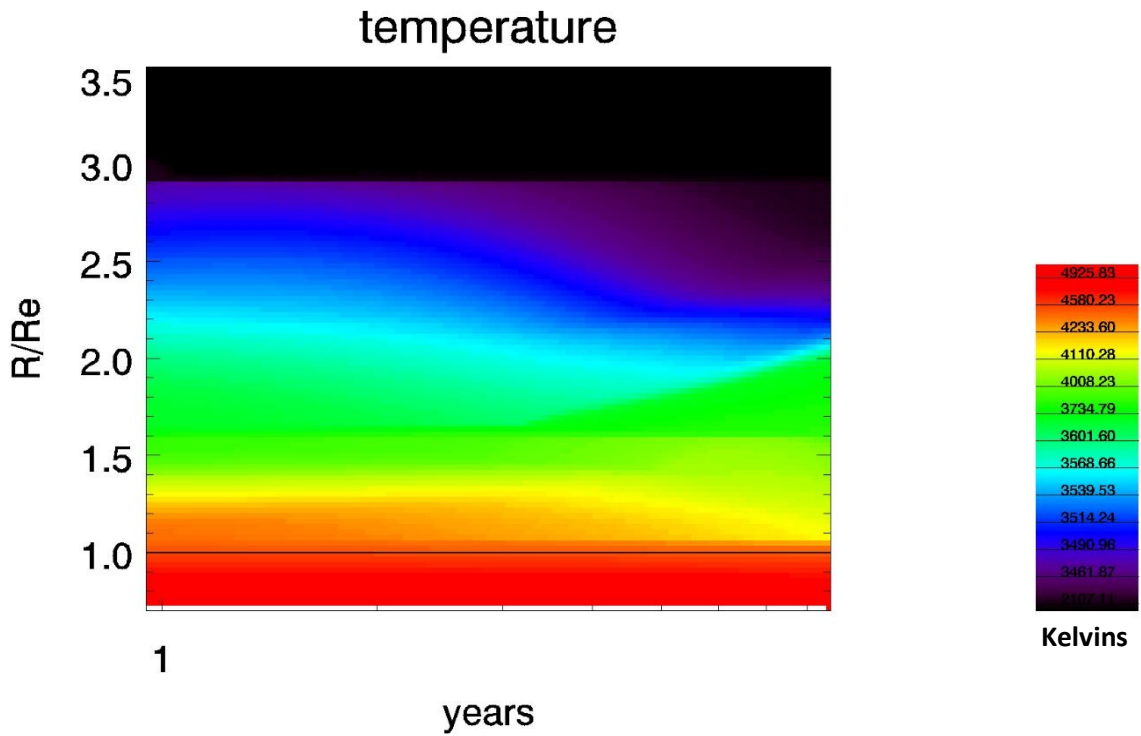

**Supplementary Figure 8:** Evolution of the disk's temperature in case a growing moon is considered at 3 Earth radii.

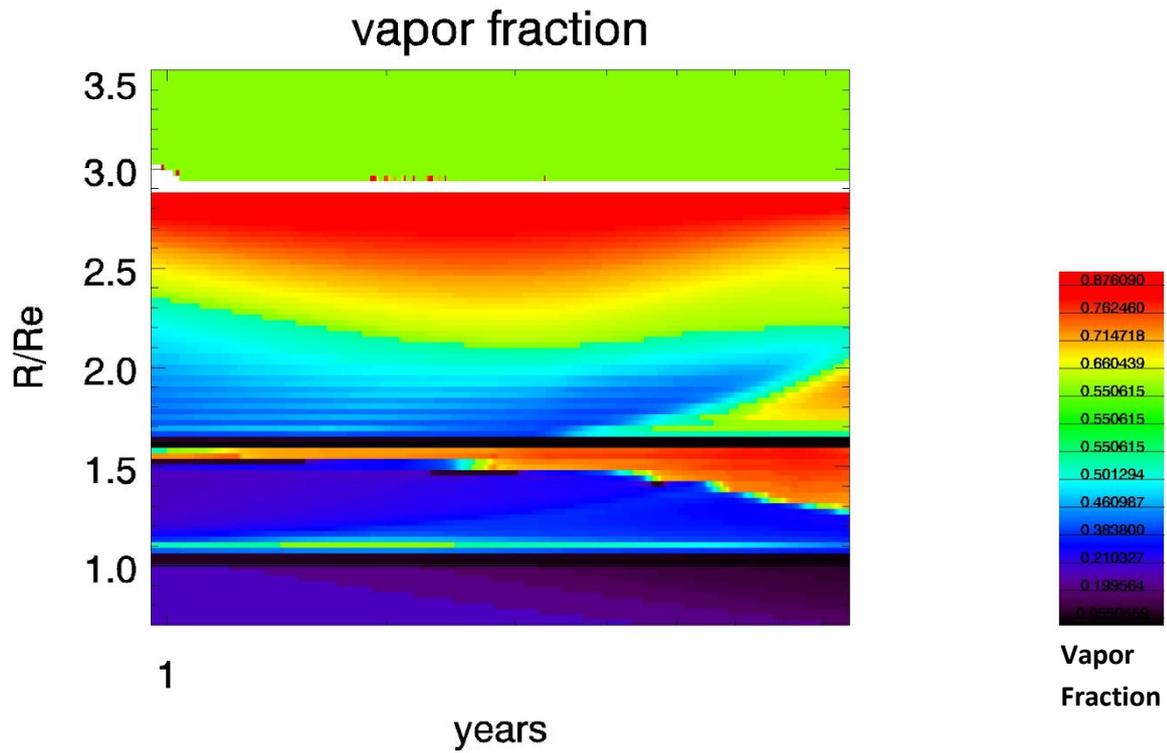

**Supplementary Figure 9**: Vapor fraction in a case of a disk with a growing moon at 3 Earth radii.

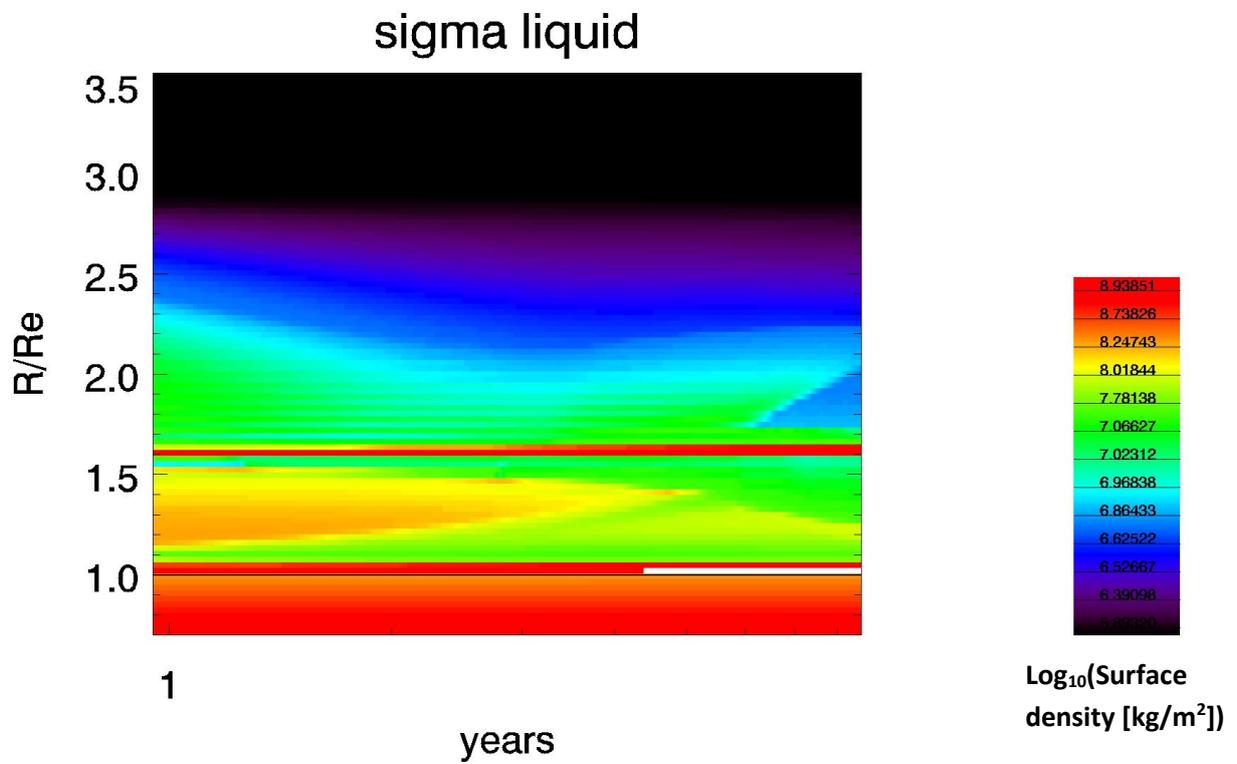

**Supplementary Figure 10:** Surface density of liquid in case of a growing moon at 3 Earth radii.